\documentclass[10pt,smallextended,apalike]{svjour3}       

\RequirePackage{fix-cm}

\smartqed  

\usepackage{graphicx} %
\usepackage{mathptmx,natbib}      
\bibpunct{(}{)}{,}{a}{}{;} 
 \journalname{J Math Biol}
\usepackage{epsfig,epic,eepic,color,graphicx,wrapfig ,amsmath,amsbsy,amssymb,amsfonts}
\usepackage{enumitem,appendix}
\usepackage{geometry} 

\usepackage{setspace,subfigure,bm,float,lscape,wrapfig}

\numberwithin{equation}{section}
\numberwithin{figure}{section}

\newcommand{\myfrac}[3][0pt]{\genfrac{}{}{}{}{#2}{\raisebox{-#1}{$#3$}}}

\newcommand{\dicty}{{\em Dictyostelium discoideum}}

\newcommand{\bdelta}{{\mbox{\bm{$\delta$}}}}
\newcommand{\bfu}{{\bm{u}}}
\newcommand{\bff}{{\mbox{\bm{$f$}}}}

\newcommand{\bfU}{{\mbox{\bm{$U$}}}}

\newcommand{\stress}{\mbox{$\mathbb{T}$}}
\def\sn{\ensuremath{\mathbf{\mathsf{ n}}}}
\def\cO{{\mathcal{O}}}
 
\newtheorem{Lemma}{Lemma}

\usepackage{lscape}

\numberwithin{equation}{section}
\numberwithin{figure}{section}

\begin{document}

\title{Computational Analysis  of Amoeboid Swimming at Low Reynolds
Number\thanks{Supported in part by NSF Grant DMS \#s 9517884 and 131974 to H. G. Othmer}
}


\author{Qixuan Wang       \and
        Hans G. Othmer   
}


\institute{Qixuan Wang \at
              540R Rowland Hall\\
              University of California, Irvine\\
              Tel.: (949) 824-3217\\
               \email{qixuanw@uci.edu}     
           \and
          Hans G. Othmer  \at
              School of Mathematics, 270A Vincent Hall \\
              University of Minnesota \\
              Tel.: (612) 624-8325\\
              Fax: (612) 626-2017\\
              \email{othmer@math.umn.edu}      
}

\date{\today}

\maketitle

\begin{abstract} Recent experimental work has shown that eukaryotic cells can
 swim in a fluid as well as crawl on a substrate. We investigate the swimming
 behavior of \dicty\, amoebae who swim by initiating traveling protrusions at
 the front that propagate rearward. In our model we prescribe the velocity at
 the surface of the swimming cell, and use techniques of complex analysis to
 develop $2$D models that enable us to study the fluid-cell interaction.  Shapes
 that approximate the protrusions used by \dicty\, can be generated via the
 Schwarz-Christoffel transformation, and the boundary-value problem that results
 for swimmers in the Stokes flow regime is then reduced to an integral equation
 on the boundary of the unit disk.  We analyze the swimming characteristics of
 several varieties of swimming \dicty\, amoebae, and discuss how the slenderness
 of the cell body and the shapes of the protrusion effect the swimming of these
 cells. The results may provide guidance in designing low Reynolds
 number swimming models.

\end{abstract}

\tableofcontents

\section{Introduction}
\label{intro}

Cell locomotion is essential throughout the development and adult forms of uni-
and multi-cellular organisms. It is beneficial in various types of taxis, in
morphogenetic movements during development, and in the immune response and wound
healing in adults, but it plays a deleterious role in cancer metastasis. When
movement is in response to extracellular signals, it involves the detection and
transduction of those signals, which can be biochemical, mechanical or of other
types, the integration of the signals into an intracellular signal, and the
spatio-temporal control of the intracellular biochemical and mechanical
responses that lead to force generation, morphological changes and directed
movement \citep{Sheetz:1999:CMF}.  Many single-celled organisms use flagella or
cilia to swim, and many mathematical models of swimming in such organisms have
been developed \citep{Lauga:2009:HSM}. The movements of eukaryotic cells that
lack such structures fall into two broad categories: mesenchymal and
amoeboid \citep{Biname:2010:WMC}.  The former can be characterized as `crawling'
in fibroblasts or `gliding' in keratocytes, and involves the extension of either
pseudopodia and/or lamellipodia driven by actin polymerization at the leading
edge.  This mode dominates in cells such as fibroblasts when moving on a 2D
substrate.  On flat surfaces the predominant protrusions are lamellipodia and
these processes suffice, but in the extracellular matrix (ECM) the protrusions
and cell body are more rounded, and the cells may also secrete matrix-degrading
proteases (MMPs) and `tunnel' their way through the ECM
\citep{Martins:2006:ECP,Mantzaris:2004:MMT}.

In the amoeboid mode cells are more rounded, and move through the ECM by
avoiding obstacles when possible, rather than removing them. Thus they rely less
on attachment to the ECM and degradation of it, and instead exploit variations
in the ECM to move through it by shape changes.  In this mode force transmission
to the extracellular matrix (ECM) depends on shape changes driven by localized
remodeling of the cytoskeleton (CSK) and myosin contraction
\citep{Insall:2009:ADL}.  Cells such as leukocytes, which normally use the
mesenchymal mode in the ECM, can migrate {\em in vivo} in the absence of
integrins, using a 'flowing and squeezing' mechanism \citep{Lammermann:2008:RLM}.
The human parasite {\em Entamoeba histolytica} uses an extreme form of amoeboid
movement called blebbing, in which the membrane detaches from the CSK
\citep{Maugis:2010:DII}, whereas zebrafish primordial germ cells and \dicty\,
(Dd) cells move using a combination of blebbing and protrusion by
spatio-temporal control of the membrane attachment to the CSK 
\citep{Blaser:2006:MZP,Diz-Munoz:2010:CDC,Yoshida:2006:DAM,Zatulovskiy:2014:BDC}.  
Cells using this mode can move up to forty times faster than those using strong
adhesion \citep{Renkawitz:2010:MFG}.  Recent experiments have shown that numerous cell
types display enormous plasticity in locomotion in that they sense the
mechanical properties of their environment and adjust the balance between the
modes by altering the balance between parallel signal transduction pathways
\citep{Renkawitz:2009:AFT,Renkawitz:2010:MFG}.  Thus crawling and swimming are
the extremes on a continuum of locomotion strategies, but many cells sense their
environment and use the most efficient strategy in a given context. 

Heretofore mathematical modeling has focused primarily on either the mesenchymal mode, in
which cells crawl via attachments to a substrate \citep{Danuser:2013:MME}, or on
microorganisms that swim using flagella or cilia
\citep{Suarez:2006:STF,Berg:1973:BSR,Lowe:1987:RRF,Gibbons:1981:CFE,Sleigh:1988:PMC,Ishimoto:2014:SES}. Here
we analyze swimming of larger cells, motivated by recent experiments which show
that both neutrophils and Dd can swim -- in the strict sense of propelling
themselves through a fluid using only fluid-cell interactions -- in response to
chemotactic gradients \citep{Barry:2010:DAN,Bae:2010:SDA,VanHaastert:2011:ACU}.
A basic question that arises is what pattern of shape changes are effective in
propelling a cell. It has been reported \citep{Salbreux:2007:SON} that freely
suspended Swiss-3T3 fibroblast cells without cell-substrate adhesion can exhibit
oscillatory shape dynamics, and some of them may also exhibit periodic bleb
dynamics correlating with the oscillations, where blebs are hemispherical
membrane blisters induced by cortical contraction
\citep{Fackler:2008:CMT,Paluch:2005:CAB}.  The combination of both dynamics may
result in a random oscillation of the cell in space.  In comparison with such
random motion, Dictyostelium and neutrophils can utilize the amoeboid swimming
mode in which the cell body is elongated and small protrusions that provide the
momentum transfer needed for motion are propagated from front to rear
\citep{Barry:2010:DAN,Bae:2010:SDA,VanHaastert:2011:ACU}.

In the following section we formulate the basic problem of swimming by shape
deformations in 2D at low Reynolds number, which is the relevant regime for
single cell movement. In 2D one can introduce a stream function, which leads to
a biharmonic equation, and the general solution of the Stokes problem is
expressed in terms of two analytic functions -- the Goursat functions -- that
are determined by the motion of the boundary of the swimmer
\citep{bouffanais2013physical,Chambrion:2011:LCS,Cherman:2000:LRN}. This in turn
leads to an integral equation for one of these functions, and the second
function can then be expressed in terms of the first. In Section~\ref{Sec.2.1}
we study the motion of Dd in 2D, and approximate the shape changes using
polygonal approximations. Using the Schwarz-Christoffel transformation, we
reduce the problem to the solution of a linear system of equations for basis
functions on the boundary of the unit disk. We show that realistic propagating
shapes can produce propulsion at speeds in the range observed experimentally
using realistic choices of the parameters.  
 
 Since cell movement, whether on a solid substrate, in a fluid, or in
a complex medium such as the ECM, involves the interplay between biochemical and
mechanical processes, dissecting the roles of each is a first step toward an
integrated description of movement.  A major  objective of our work is to answer a question
posed by experimentalists, which is `How does deformation of the cell body
 translate into locomotion?' \citep{Renkawitz:2010:MFG}.  A longer-range goal is to produce a unified
 description for swimming that integrates signaling and mechanics in viscous and
 viscoelastic environments similar to the extracellular tissue environment.  As
 one experimentalist stated 'the complexity of cell motility and its regulation,
 combined with our increasing molecular insight into mechanisms, cries out for a
 more inclusive and holistic approach, using systems biology or computational
 modeling, to connect the pathways to overall cell behavior'
 \citep{Insall:2009:ADL}.

\section{Low Reynolds Number Swimming Problems}
\label{Sec.1}

\subsection{General description of swimming mechanics at low Reynolds numbers}
\label{Sec.1.1}
 
 Recent interest in the motion of biological organisms in a viscous fluid was
 re-kindled by Purcell's 1977 description of life at low Reynolds number (LRN) 
 \citep{Purcell:1977:LLR}, and a wide variety of applications have been analyzed
 since then.  A review of some of these is given elsewhere
 \citep{Lauga:2009:HSM}, and we only describe the relevant background.  We
 consider motion in an incompressible Newtonian fluid of density $\rho$,
 viscosity $\mu$, and velocity $\bfu$, for which the governing equations are
\begin{align}
\label{NSeqn}
\rho\dfrac{\partial \bfu}{\partial t} +\rho (\bfu \cdot\nabla) \bfu &=  \nabla
 \cdot  \stress  + \bff   
 =- \nabla p  + \mu \Delta \bfu + \bff, \\
   \nabla \cdot \bfu &= 0 
\end{align}
where $\stress = -p \bdelta + \mu (\nabla \bfu + (\nabla \bfu)^T)$ is the
viscous  stress tensor.  ${\bf f}$ is the external force field, which we assume to
be zero or include it in the pressure hereafter. This is appropriate for the experimental configuration
described in \citep{Barry:2010:DAN}, since the fluid  had a variable density
that allowed cells to find a point of neutral buoyancy. The Reynolds number
based on a characteristic length scale $L$ and speed scale $U$ is Re = $\rho LU
/\mu$, and when converted to dimensionless form and the symbols re-defined, the
equations read
\begin{eqnarray}
\label{NSeqn2}
Re\cdot \!Sl\dfrac{\partial \bfu}{\partial t} + Re (\bfu \cdot\nabla) \bfu &=& - \nabla p
  +  \Delta \bfu \\
  \nabla \cdot \bfu &=& 0. \nonumber 
\end{eqnarray}
Here $Sl = \omega L/U $ is the Strouhal number and $\omega$ is a characteristic
frequency of the shape changes.  When Re$\,\ll1$ the convective momentum term in
(\ref{NSeqn2}) can be neglected, but the time variation requires that
$Re\cdot\!Sl \equiv \omega L^2/\nu \ll 1$. This is not always true, even for
small swimmers \citep{ishimoto2013spherical,wang2012unsteady}, but as we show
next, it applies here.

The small size and slow speed of cells considered here leads to LRN flows, and
 in this regime cells move by exploiting the viscous resistance of the fluid.
 For example, Dd amoebae have a typical length $L \sim 25
\mu \textrm{m} $ and can swim at $U \sim 3 \mu \textrm{m} / \textrm{min}$
\citep{VanHaastert:2011:ACU}. Assuming the medium is water $(\rho \sim 10^3
\textrm{kg} \ \textrm{m}^{-3}, \ \mu \sim 10^{-3}\, \textrm{Pa-s})$, 
and the deformation frequency $\omega \sim 1 /s$, $Re \sim \cO (10^{-6})$ and
$Sl \sim \cO (10^{-4})$.  In fact the experiments are done in oil that
is significantly more viscous \citep{Barry:2010:DAN}.  In any case, when both inertial terms on the
left hand side of (\ref{NSeqn2}) are neglected, which we
assume hereafter, the flow is governed by the Stokes equations
\begin{equation}
\mu \Delta \bfu  - \nabla p   =  {\bf 0},  \qquad \qquad \nabla \cdot \bfu = 0. 
\label{creep}
\end{equation}
 Throughout we consider the propulsion problem in an infinite domain and assume
 that the fluid is at rest at infinity. 

 Since time does not appear in the equations, a time-reversible stroke produces
 no net motion, which is the content of the famous `scallop theorem'
 \citep{Purcell:1977:LLR}. Furthermore, since we assume that there is no inertia
 in the fluid (momentum transfer is assumed to be instantaneous), in the Stokes
 regime there is no net force or torque on a self-propelled swimmer moving in an
 infinite fluid that is at rest at infinity. One can see this by integrating the
 stress over the boundary of the swimmer and then equating this via the momentum
 equation to the stress on a circle at infinity
 \citep{Shapere:1989:GSP}. Therefore movement is a purely geometric process: the
 net displacement of a swimmer during a stroke is independent of the rate at
 which the stroke is executed when $Re =0$, and approximately so as long as
 $Re\cdot\!Sl $ remains small enough.

 Let $\Omega(t) \in R^n$ be the swimmer (a compact set with a sufficiently
 smooth boundary). Throughout we consider a fixed global reference frame
 $(\bm{x},t)$ and a body frame $(\bm{X},t)$ attached to the swimmer, where time
 is scaled the same in both frames.  We use the notation $\partial
 \Omega_{\bm{x}} (t)$ and $\partial \Omega_{\bm{X}} (t)$ to denote the boundary
 of $\Omega$ in the fixed and body frames, resp., and when observed from the
 body frame, the shape deformations of the swimmer are specified as
 $\bm{u}_{\textrm{s}} (\bm{X}, t)$ for $\bm{X} \in \partial\Omega_{\bm{X}} (t)$.
 Denote the uniform (rigid-body) translational and rotational velocities of the
 swimmer when observed from the fixed frame as $\bm{U} (t)$ and $\bm{\omega }
 (t)$, respectively.  Then the instantaneous velocity on the swimmer's surface
 $\partial\Omega_{\bm{x}} (t)$ is
\begin{eqnarray}
\label{eq.intro.7}
 \bm{u} (\bm{x},t)  = \bm{U} (t) + \bm{\omega} (t) \times \bm{x} + \bm{u}_{\textrm{s}}  (\bm{X},t)
\end{eqnarray}
where  $\bm{u}$  satisfies ($\ref{creep}$).

A \textit{swimming stroke} is specified by a time-dependent sequence of shapes,
 and it is \textit{cyclic } if the initial and final shapes are identical
 \citep{Shapere:1989:GSP}. The canonical LRN self-propulsion problem is -- {\em
 given a cyclic shape deformation specified by $\bm{u}_s$, solve the Stokes
 equations subject to the zero force and torque conditions }
\begin{equation}
\label{eq2}
\bm{F}(t) \equiv  \int_{\partial\Omega_{\bm{x}}(t)} \mathbb{T} \cdot \sn = 0, \qquad
 \bm{T} (t)  \equiv \int_{\partial\Omega_{\bm{x}} (t)}{\bm{x}}\wedge(  \mathbb{T}  \cdot \sn) = 0
\end{equation}
{\em and the boundary conditions  
\begin{equation}
\label{eq2a}
   \bm{u} |_{\partial\Omega} = \bm{u}_{\textrm{s}}  + \bfU +   \bm{\omega}  \times \bm{x} , \qquad
\bm{u}|_{\bm{x} \rightarrow \infty} =\bm{0}  
\end{equation}
where $\sn$ is the exterior normal.  }

How does the swimmer move if force- and torque-free?  It would appear that at
best it can only translate or rotate at a constant rate. However, an arbitrary
deformation of the surface will not satisfy (\ref{eq2}) in general, and in
particular, will generate a flow at infinity. However, since we assume that the
fluid is at rest there this flow must be counteracted by an imposed flow, which
then defines the rigid linear and angular velocities of the swimmer, and leads
to satisfaction of the force-free and torque-free conditions \citep{Shapere:1989:GSP}. Since the
shape changes are time-dependent the counterflow is also, as are the rigid
translations and rotations.

Thus the canonical LRN swimming problem is: {\em 
given a cyclic sequence of shape deformations by specifying $\bm{u}_{\textrm{s}}
$ on the boundary $\partial \Omega_{\bm{X}}$ of the swimmer, solve the Stokes
equations~($\ref{creep}$) for a trial  velocity field $\bm{u}$  and then use  the
force-free and torque-free conditions to determine $\bm{U}$ and
$\bm{\omega}$ so as to satisfy the boundary conditions at infinity.} In what
follows we restrict attention to swimming in two space dimensions, and describe
 the problem as the 2D LRN swimming (2DLRNS) problem.

 \subsection{The 2D swimmer} 
\label{Sec.1.2}

 In two space dimensions  techniques from complex analysis can be used to significantly simplify
the problem of computing solutions to the Stokes equations for LRN swimming
problems.  Two main methods have been developed that can be applied, one that
was first developed by Muskhelishvili to solve problems in elasticity
\citep{Muskhelishvili:1977:SBP}, and another that is essentially a boundary
integral method\citep{Pozrikidis:1992:BIS} for 2D problems
\citep{Greengard:1996:IEM,Kropinski:1999:IEM,Kropinski:2001:ENM,Kropinski:2002:NMM,Kropinski:2011:ENM}.
 Significant analytical insights can be gained using Muskhelishvili's method,
 including the application of  control theory to LRN  swimming
\citep{Shapere:1989:GSP,Shapere:1989:ESP,Kelly:1998:MCR,Kelly:2000:MEP}, and we
use this  method in this paper.  

In $2D$ the incompressibility condition $\nabla \cdot \bm{u} = 0$ in
($\ref{creep}$) can be satisfied by introducing a stream function
$\Lambda (z, \overline{z}; t)$,  which is a real-valued scalar potential such that 
\begin{eqnarray*}
u =   \dfrac{\partial \Lambda}{\partial y} - i \dfrac{\partial \Lambda}{\partial x}.  
\end{eqnarray*}
Here and hereafter we use $u \in \mathbb{C}$ to denote the velocity field
(denoted by $\bm{u}$ in Section \ref{Sec.1.1}) in the complex
$z$-plane.  Then the Stokes equations~(\ref{creep}) imply that $\Lambda $
satisfies the  biharmonic equation
\begin{eqnarray}
\label{eq.review2D.3}
 \Delta^2 \Lambda = 0.
\end{eqnarray}
The general solution of ($\ref{eq.review2D.3}$) can be expressed by
Goursat's formula \citep{Muskhelishvili:1977:SBP}
\begin{eqnarray}
\label{eq.review2D.4}
 \Lambda (z, \overline{z} ; t) = \Re \big[ \bar{z} \phi (z ; t ) + \chi (z; t ) \big]
\end{eqnarray}
where for any $t$, $ \phi (z; t)$ and $ \chi (z ;t )$ are analytic functions on
the fluid domain $\mathbb{C} \setminus \Omega (t)$ and continuous on $\mathbb{C} \setminus
\textrm{int} \Omega (t)$, where $\textrm{int}\, \Omega $ denotes the interior of
$\Omega$.  $ \phi (z ;t )$ and $ \chi (z ;t )$ are the
\textit{\textbf{Goursat functions}}. To simplify the expression of physical
quantities and the  following discussion, we hereafter impose the substitution 
 $\phi \rightarrow - i \phi$, $\chi \rightarrow - i \chi$,  as suggested 
by Shapere and Wilczek \citep{Shapere:1989:GSP}, and Table \ref{tab.1} gives the expressions
of several physical quantities in terms of these functions. In the table and
hereafter we denote $\partial_z$ by $'$ for simplicity,  $n = - i dz/ds$
gives the exterior normal to $\partial \Omega $ ({\em i.e.,}  directed into the fluid domain),
and $s$ denotes the arc length, traversed counterclockwise.
\begin{table}[htbp]
\center
\caption{Representation of various  physical quantities by the Goursat functions. }

\label{tab.1}       
%
\begin{tabular}{|l|l|}
\hline  
 & \\
Velocity &  $u = \phi (z) - z \overline{\phi ' (z)} - \overline{\chi ' (z)} $   \\
 & \\
 \hline 
  & \\
Pressure & $p = - 4 \mu \Re \{ \phi ' (z) \} $ \\
 & \\
\hline
 & \\
Vorticity & $\vartheta = - 4   \Im \{ \phi' (z) \} $ \\ 
 & \\
\hline
 & \\
Stress force & $ f = 4 \mu \Re (\phi') n - 2 \mu (z \overline{\phi''} + \overline{\chi''} ) \overline{n}$ \\
 & \\
 \hline
 & \\
Stress force &  $f d s = - 2 i \mu d \big( \phi + z \overline{\phi'} + \overline{\chi'} \big)
$ \\
(differential form) & \\
 \hline
\end{tabular}
\end{table}

For swimming problems in a $2D$ Stokes flow in the unbounded domain $\mathbb{C} / \Omega (t)$, 
we require that the stress vanish at infinity, and as a result the Goursat functions
must take the general form \citep{Muskhelishvili:1977:SBP,Greengard:1996:IEM}:
\begin{eqnarray}
\label{eq.review2D.10}
 \phi (z,t) &=& - \dfrac{X (t) + i Y (t)}{2 \pi (1 + \kappa (t))} \log z  + \widetilde{\phi} (z,t)   \\   
\label{eq.review2D.11}
\psi(z,t) = \chi' (z,t) &=& \dfrac{X (t)- i Y (t)}{2 \pi (1 + \kappa (t))} \log z +  \widetilde{\psi} (z; t)
\end{eqnarray}
where $\widetilde{\phi} (z,t) $ and $\widetilde{\psi} (z,t)$ are single-valued
and analytic on $\overline{\mathbb{C}} / \Omega$ (where $\overline{\mathbb{C}} =
\mathbb{C} \cup \{ \infty \}$).  For the self-propulsion problem we must also
require that $X = Y = 0$ to ensure a bounded velocity at infinity. We then
compute the translational and rotational velocity for a trial pair
$(\widetilde{\phi},\widetilde{\chi})$, and when these are subtracted from the
flow the motion is force-free and torque-free
\citep{Muskhelishvili:1977:SBP,Greengard:1996:IEM}. Thus for $2$D swimming
problems, the Goursat functions $\phi (z)$ and $\psi (z)$ should be
single-valued and analytic on $\overline{\mathbb{C}} / \Omega$, and therefore  they
 have Laurent expansions in $\overline{\mathbb{C}} / \Omega$ of the following form. 
\begin{eqnarray}
\label{eq.review2D.25}
 \phi (z,t) &=& a_0 (t) + \dfrac{a_{-1} (t)}{z} + \dfrac{a_{-2} (t)}{z^2} +
 \cdots \\   
\label{eq.review2D.26}
 \psi (z,t) &=& b_0 (t) + \dfrac{b_{-1} (t) }{z} + \dfrac{b_{-2} (t) }{z^2} + \cdots
\end{eqnarray}

As a result, the 2DLRNS problem with specified shape changes as described above
can be rephrased as follows -- {\em find functions $\phi(z,t)$ and $\psi (z,t)$
that are analytic on $\overline{\mathbb{C}} / \Omega (t) $ and continuous on
$\overline{\mathbb{C}} / \textrm{int}\; \Omega (t)$ such that}
\begin{eqnarray}
\label{eq.review2D.12}
 \phi (z, t) - z \overline{\phi' (z; t)} - \overline{\psi (z, t)} = V (z, \overline{z}, t) \qquad (z \in \partial \Omega (t)).
\end{eqnarray}
Here $V (z, \overline{z}, t) $ for $z \in \partial \Omega (t)$ is the velocity
boundary condition determined by the shape changes, namely, the complex form of
$\bm{u}_{\textrm{s}}$ as introduced in
Section~$\ref{Sec.1.1}$. Equation~($\ref{eq.review2D.12}$) will be referred to as
\textit{the boundary condition constraint } on $\phi$ and $\psi$ in the study of 2D LRN swimming.

\subsection{Pull-back of the problem to the disk  and derivation of the Fredholm integral equation} 
\label{Sec.1.3}

As we mentioned earlier, there are two main methods for solving the 2D LRN
problem -- which is to say to solve ($\ref{eq.review2D.12}$) -- Muskhelishvili's
method and the integral equation method. In the integral equation method the
velocity boundary condition leads to an integral operator defined on the
swimmer's boundary, while in Muskhelishvili's method the approach is to first
map the $z$-plane to a fixed complex computational $\zeta$-plane, on which the
integral operator is applied. In the first case one deals with a moving boundary
whose stress field is specified, while in Muskhelishvili's method we consider
velocity boundary condition and the problem can be pulled back into a fixed
boundary problem. Generally speaking, the integral equation method is useful
when the stress field along the boundary is known or multiple bodies are
involved. On the other hand, for a single deformable swimmer it is easier to
treat a sequence of shape deformations using Muskhelishvili's method, which also
facilitates the use of control theory to 2D LRN swimming systems and simplifies
the design and study of micro aquatic robots.  Hereafter we focus on the use of
Muskhelishvili's method.

Suppose that the cell occupies an open, simply-connected, bounded region $\Omega
(t)$ in the complex $z$-plane at time $t$. Let $D = \{ \zeta \in \mathbb{C}: |
\zeta | < 1 \}$ be the unit disk in the computational $\zeta$-plane. The regions
exterior to $\overline{D}$ and to $ \overline{\Omega} (t) $ in the extended
complex planes, i.e., $\overline{\mathbb{C}} / \overline{D}$ and
$\overline{\mathbb{C}} / \overline{\Omega} (t) $, are both simply-connected as
the infinity point $\infty$ is included in $\overline{\mathbb{C}} $.  The
\textit{Riemann mapping theorem} \citep{Ahlfors:1978:CA} ensures the existence of
a single-valued analytic conformal mapping $z = w (\zeta; t)$ which maps
$\overline{\mathbb{C}} / \overline{D}$ one-to-one and onto
$\overline{\mathbb{C}} /\overline{\Omega} (t) $, and preserves the
correspondence of infinity, i.e., $w (\infty; t) \equiv \infty$.  Moreover, this
mapping can be extended continuously to $\overline{\mathbb{C}} / D$
\citep{Younes:2010:SD}, and $w$ maps $\partial (\overline{ \mathbb{C}} / D) =
S^1$ to $\partial\Omega (t)$.
%
\begin{figure}[htbp]
\centering
\includegraphics[width=0.8\textwidth]{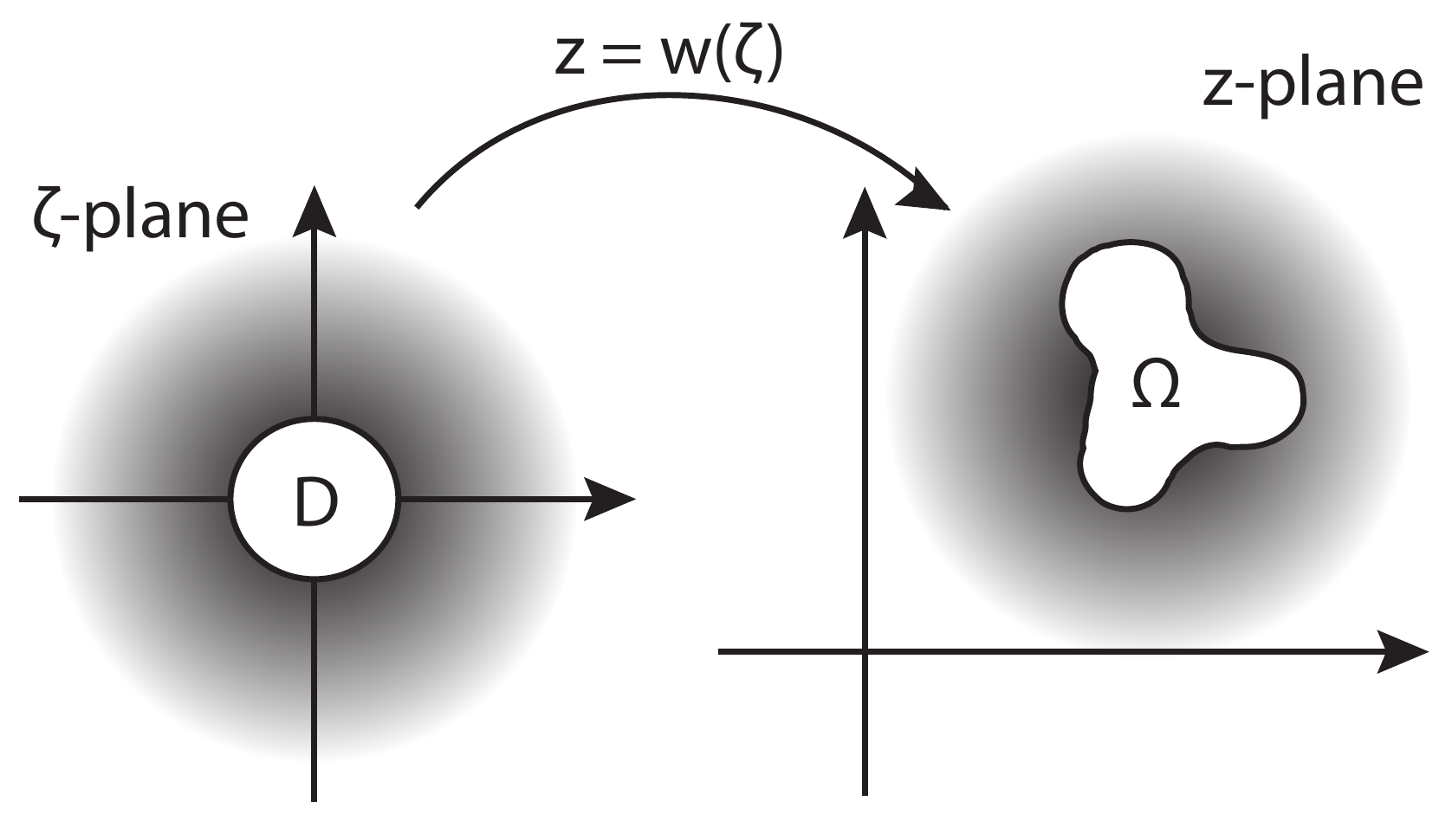}
\caption[The conformal mapping of the fluid domain]{The conformal mapping from
$\overline{\mathbb{C}} / D $ to $\overline{\mathbb{C}} / \Omega $.} 
\label{fig.review2D.1}
\end{figure} 

We assume that the mapping preserves the point at infinity, and then have the
following result regarding the form of its Laurent expansion
\citep{Ahlfors:1978:CA}.
\begin{Lemma} 
Suppose that $\Omega (t)$ is a non-empty open bounded simply-connected domain in
$\mathbb{C}$, and that $z = w (\zeta ; t)$ is a conformal mapping from the
exterior of the unit disk $D$ to $  \Omega^c$ that preserves the
point at infinity. Then $w (\zeta;t)$ has a Laurent expansion of the form
\begin{eqnarray}
\label{eq.review2D.13}
w (\zeta;t) = \alpha_{ 1} (t) \zeta + \alpha_0 (t) + \dfrac{\alpha_{-1} (t)}{\zeta} + \dfrac{\alpha_{-2} (t)}{\zeta^2} 
+ \cdots \dfrac{\alpha_{-n} (t)}{\zeta^n} + \cdots  
\end{eqnarray}
where $\alpha_{ 1} (t) \neq 0 \ \textrm{and} \ |\zeta| > 1$.
\label{Prop.2}
\end{Lemma}
When we consider in the body frame, $\Omega(t)$ gives the configuration of the swimmer,
in which case we require $\alpha_0 = 0 $ and $\alpha_{-1} \in \mathbb{R}$ in ($\ref{eq.review2D.13}$).

Hereafter  $w (\zeta;t)$ is the conformal map which maps
$\overline{\mathbb{C}}/ \overline{D}$ onto $\overline{\mathbb{C}} / \Omega (t) $
such that  $w (\infty; t) \equiv \infty$, extended to  the boundary of the swimmer,
which is given by $\partial\Omega (t) = \{ z (t) = w (\sigma; t); \sigma \in S^1 \}$.
We impose a no-slip  condition on the boundary, and therefore
\begin{eqnarray}
\label{eq.review2D.14}
 u \big( w (\sigma) \big) (t) = \dfrac{\partial}{\partial t } w (\sigma; t).
\end{eqnarray}

Suppose that $\phi (z,t)$ and $\psi (z,t)$ are the solutions to the boundary
condition constraint given by ($\ref{eq.review2D.12}$). Let
\begin{eqnarray*}
 \Phi (\zeta;t) = \phi (w (\zeta;t);t), \qquad \Psi (\zeta) = \psi (w (\zeta;t);t) \qquad (|\zeta| \geq 1)
\end{eqnarray*}
thus $ \Phi (\zeta;t)$ and $\Psi (\zeta;t)$ are functions on $\zeta$-plane which are 
analytic on $\mathbb{C} / \overline{D}$ and continuous on $\mathbb{C} / D$ at any time $t$. Hence on 
$|\zeta| \geqslant 1$, we have  Laurent expansion for $\Phi$ and $\Psi$ given
by 
\begin{eqnarray}
\label{eq.review2D.15}
 \Phi (\zeta;t) &=& A_0 (t) + \dfrac{A_{-1} (t)}{\zeta} + \dfrac{A_{-2}
 (t)}{\zeta^2} + \cdots  \\   
\label{eq.review2D.16}
 \Psi (\zeta;t) &=& B_0 (t) + \dfrac{B_{-1} (t)}{\zeta} + \dfrac{B_{-2} (t)}{\zeta^2} + \cdots 
\end{eqnarray}
where $A_k, B_k$ ($k=0, -1, -2, \cdots$) are continuous functions from $[0, \infty] $ to $\mathbb{C}$.

If we substitute these expressions into ($\ref{eq.review2D.12}$) and omit the time index $t$, 
then the velocity field can be expressed as a function on the $\zeta$-plane as
\begin{eqnarray}
\label{eq.review2D.17}
\widetilde{u}  (\zeta, \overline{\zeta}) : = u \big( w (\zeta), \overline{w (\zeta)} \big)
= \Phi (\zeta) - \dfrac{w (\zeta)}{\overline{w' (\zeta)}} \overline{\Phi' (\zeta)} - \overline{\Psi (\zeta)}
\end{eqnarray}
where $'$ represents $\partial_\zeta$. If we let $V (\sigma ; t)$ denote the
boundary condition of the velocity in general, then the boundary condition
constraint ($\ref{eq.review2D.12}$) can be generalized to the condition
\begin{eqnarray}
\label{eq.review2D.18}
\Phi (\sigma) - \dfrac{w (\sigma)}{\overline{w' (\sigma)}} \overline{ \Phi' (\sigma) } - \overline{\Psi ( \sigma)} =  V (\sigma)
\qquad (\sigma \in S^1).
\end{eqnarray}
In the 2DLRNS problem the boundary condition is given as in ($\ref{eq.review2D.14}$),
{\em i.e.}, $ V (\sigma ; t) =  \partial_t w (\sigma;t)$, but in the discussion below  we 
will adopt the  more general form  $V (\sigma ;t) = \sum_{n } \lambda_n (t) \sigma^{ n}$.

Equation (\ref{eq.review2D.18}) involves both $\Phi$ and $\Psi$, but this can be
reduced to an integral equation in $\Phi$ alone using a technique developed in
\cite{Muskhelishvili:1977:SBP}, where it was originally developed for the
solution of problems in elasticity rather than the Stokes flows, and the
resulting integral equation is the same. We present the precise statement here
for completeness.  The reduction relies on a decomposition of the boundary
condition $V (\sigma)$ derived from the following Plemelj formula
\citep{sokhotskii1873definite,Cima:2006:CT,England:2012:CVM}. 

\newtheorem{Theorem}{Theorem} 
\begin{Theorem}[Plemelj Formula] 
Suppose $V (\sigma)$ is continuous on $S^1 = \{ \sigma \in \mathbb{C}; |\sigma| = 1 \}$ and, 
for a particular $\sigma_0 \in S^1$, satisfies the H\"{o}lder condition
\begin{eqnarray}\label{eq.review2D.19}
 \big| V (\sigma) - V (\sigma_0) \big| \leq C \big| \sigma - \sigma_0 \big|^\alpha , \qquad \sigma \in S^1
\end{eqnarray}
for some positive constants $C$ and $\alpha$. Let
\begin{eqnarray}\label{eq.review2D.20}
  \widehat{V} (\zeta) = \dfrac{1}{2 \pi i} \int_{S^1} \dfrac{V (\sigma)}{\sigma - \zeta} d \sigma
\end{eqnarray}
for $\zeta \in \mathbb{C}/S^1$. Then the limits
\begin{eqnarray}\label{eq.review2D.21}
 V^- (\sigma_0) := \lim_{r \rightarrow 1^-} \widehat{V} \big(r \sigma_0 \big) \qquad \textrm{and} \qquad 
V^+ (\sigma_0) := \lim_{r \rightarrow 1^-} \widehat{V} (\sigma_0/r) 
\end{eqnarray}
exist and moreover, $V^- (\sigma_0) - V^+ (\sigma_0) = V (\sigma_0)$. Furthermore, the Cauchy principal-value integral
\begin{eqnarray*}
 P.V. \int_{S^1} \dfrac{V (\sigma)}{\sigma - \sigma_0} d \sigma := 
\lim_{\varepsilon \rightarrow 0} \int_{|\sigma - \sigma_0| > \varepsilon} 
 \dfrac{V (\sigma)}{\sigma - \sigma_0} d \sigma
\end{eqnarray*}
exists and
\begin{eqnarray*}
 V^+ (\sigma_0) + V^- (\sigma_0) = \dfrac{1}{\pi i} P.V. \int_{S^1} \dfrac{V (\sigma)}{\sigma - \sigma_0} d \sigma
\end{eqnarray*}
\label{Thm.2}
\end{Theorem}
We assume that the boundary condition $V (\sigma)$ in the swimming problems 
satisfies the H\"{o}lder condition as in  ($\ref{eq.review2D.19}$) for any $\sigma_0 \in S^1$. 
For the function $ \widehat{V} (\zeta)$ defined in  ($\ref{eq.review2D.20}$), let 
$V^+ (\zeta) =   \widehat{V} (\zeta)$ for $|\zeta| > 1$, and 
$V^- (\zeta) =   \widehat{V} (\zeta)$ for $|\zeta| < 1$, then
$V^- (\zeta)$ is analytic for $|\zeta| < 1$, while $V^+ (\zeta)$
is analytic for $|\zeta| > 1$ with $V^+ (\zeta) \rightarrow 0$ as $\zeta \rightarrow \infty$.
Moreover, according to the Plemelj formula,
both $V^+$ and $V^-$ can be continuously extended to $S^1$ by  ($\ref{eq.review2D.21}$), 
and we have the decomposition $V = V^- - V^+$ on $S^1$. 

To continue, we first introduce some notation. For an analytic function
$f(\zeta)$, we define $\overline{f} (\zeta)$ as $\overline{f} (\zeta) =
\overline{f (\overline{\zeta}) }$ \citep{Muskhelishvili:1977:SBP}.  With this
notation, suppose that $f (\zeta)$ is analytic on $|\zeta| \gtrless R$ for some
$R >0$, then $\overline{f} (1 / \zeta)$ is analytic on $|\zeta| \lessgtr R$.
The proof of this is straightforward.  Without loss of generality we assume that
$f (\zeta)$ is analytic on $|\zeta| > R$, and then on $|\zeta| > R$ we have the
following Laurent expansion for $f (\zeta)$
\begin{eqnarray*}
 f (\zeta) = f_0 + \dfrac{f_1}{\zeta} + \dfrac{f_2}{\zeta^2 } + 
\cdots + \dfrac{f_n}{\zeta^n} + \cdots
\end{eqnarray*}
Hence 
\begin{eqnarray*}
 f \big( \dfrac{1}{\zeta} \big) = f_0 +  f_1 \zeta  +  f_2 \zeta^2   + 
\cdots +  f_n \zeta^n + \cdots
\end{eqnarray*}
which is easily seen to be analytic on $|\zeta| < R$.
By definition of the function $\overline{f}$, we have
\begin{eqnarray*}
 \overline{f} \big( \dfrac{1}{\zeta} \big)
= \overline{f  \big( \dfrac{1}{\overline{\zeta}} \big)} 
=\overline{f_0 +  f_1 \overline{\zeta}  +  f_2 \overline{\zeta}^2   + 
\cdots +  f_n \overline{\zeta}^n + \cdots}
= \overline{f_0} + \overline{ f_1} \zeta  +  \overline{ f_2} \zeta^2   + 
\cdots +  \overline{ f_n} \zeta^n + \cdots
\end{eqnarray*}
which is easily seen to be analytic on $|\zeta| < R$ as well. This establishes
the assertion.

To derive an integral equation for $\Phi$, let  $|\zeta| > 1$ and  apply the functional operator 
\begin{eqnarray*}
\dfrac{1 }{2 \pi i} \int_{S^1} \dfrac{\bullet}{\sigma - \zeta} d \sigma
\end{eqnarray*}
to both sides of  ($\ref{eq.review2D.18}$). $\Phi (\zeta)$ is analytic on $|\zeta| > 1$,
and continuous on $|\zeta| \geq 1$, so
\begin{eqnarray*}
\dfrac{1}{2 \pi i} \int_{S^1} \dfrac{\Phi (\sigma)}{ \sigma - \zeta} d \sigma= - \Phi (\zeta)
\end{eqnarray*}
for $|\zeta| > 1$.
Moreover, $\overline{\Psi} (1/\zeta)$ is analytic on $|\zeta| < 1$ and continuous on $|\zeta| \leq 1$, and thus
\begin{eqnarray*}
\dfrac{1}{2 \pi i} \int_{S^1} \dfrac{\overline{\Psi (\sigma)}}{\sigma - \zeta} d \sigma= 
\dfrac{1}{2 \pi i} \int_{S^1} \dfrac{\overline{\Psi} \big( 1 / \sigma \big)}{\sigma - \zeta} d \sigma= 0
\end{eqnarray*}
On the other hand, for the right-hand side of  ($\ref{eq.review2D.18}$) 
by Plemelj formula we have
\begin{eqnarray*}
\dfrac{1}{2 \pi i}  \int_{S^1} \dfrac{V (\sigma)}{ \sigma - \zeta} d \sigma = V^+ (\zeta)
\end{eqnarray*}
Thus  the final result is  Fredholm integral equation
\begin{eqnarray}
\label{eq.review2D.22}
 \Phi (\zeta) + \dfrac{1}{2 \pi i} \int_{S^1} \dfrac{w (\sigma)}{\overline{w' (\sigma)}} \dfrac{\overline{\Phi' (\sigma)}}{\sigma - \zeta} d \sigma = - V^+ (\zeta)
\qquad ( |\zeta| \geq 1)
\end{eqnarray}
 where $- V^+ (\zeta)$ is the analytic part of $V (\zeta)$ when $|\zeta| > 1$,
namely, $- V^+ (\sigma ;t) = \sum_{n \leq 0 } \lambda_n (t) \sigma^{ n}$.  This
equation has only one unknown function $ \Phi$ and once it is known $\Psi$ can
be obtained from (\ref{eq.review2D.18}). The joint solution is unique in the
sense that the constant terms $A_0$ and $B_0$ of $\Phi$ and $\Psi$ in
equations~($\ref{eq.review2D.15}$,$\ref{eq.review2D.16}$) may vary, but the
difference $A_0 - \overline{B_0}$ is uniquely determined.

\subsection{Expressions of physical quantities by the pull-back of Goursat
functions}
\label{Sec.1.4}
  
Given  the pull-back of the Goursat functions determined by
 $(\ref{eq.review2D.22})$,  whose Laurent expansions are in the forms
given in  equations~($\ref{eq.review2D.15}, \ref{eq.review2D.16}$), we can obtain the
expressions of several physical quantities of  interest in  a
typical swimming problem
\citep{Shapere:1989:GSP,Cherman:2000:LRN,Shapere:1989:ESP,Avron:2004:OSL}.  In
the following discussion we scale the length by $R$ and the time by $T$, where
$R$ usually corresponds to the radius of the cell when it is in the shape of  a disk,
{\em i.e.}, $\pi R^2 = \textrm{Area of the cell}$.

\begin{enumerate}
 \item \textit{Rigid motions (I): translation}. 
Following the approach ussed in \citep{Shapere:1989:GSP}, let 
$U_\infty$ and $\omega_\infty$ be the translational and rotational components of the far field 
behavior when prescribing boundary condition~($\ref{eq.review2D.14}$), respectively; then the actual translational velocity $U$ and the rotational velocity $\omega$
of the swimmer are given by
\begin{eqnarray}\label{RigidMotionvsInfty}
U = - U_\infty, \qquad \omega = - \omega_\infty
\end{eqnarray}
as the fluid is static at far field.
As shown in appendix~$\ref{Appendix:A_tr}$, $U_\infty$ can be computed from the relation 
\begin{eqnarray}
\label{eq.review2D.24}
 U_\infty = a_0 - \overline{b_0} = A_0 - \overline{B_0}   
\end{eqnarray}
where the $a_0$, $b_0$ are the leading order terms of $\phi$ and $\psi$ as given
in equations~($\ref{eq.review2D.25}$,$\ref{eq.review2D.26}$), and $A_0$, $B_0$
are the leading order terms of $\varPhi$ and $\Psi$ as given in
equations~($\ref{eq.review2D.15}$,$\ref{eq.review2D.16}$).  Then the net
translation of the swimmer at any instant $t$ is given by
\begin{eqnarray*}
 Tr (t) = \int_0^t - U_\infty (t) \ dt = \int_0^t \Big[ - A_0 (t) + \overline{ B_0 (t)} \Big] \ dt
\end{eqnarray*}
and therefore the average velocity of the swimmer within a period is
$\widetilde{U}  = Tr (T) / T$.

 \item \textit{Rigid motions (II): rotation}. 
The rotational velocity $\omega$ can be obtained from the following
relations. 
\begin{eqnarray}
\label{eq.review2D.27}
 \omega = \dfrac{T (V ; w)}{T (V^{\textrm{rot}}; w)} =
 \dfrac{ - 4 \pi \mu \Im b_{-1} }{T (V^{\textrm{rot}}; w)}
 =\dfrac{ - 4 \pi \mu \Im \big( B_{-1} \alpha_{ 1} \big)}{T (V^{\textrm{rot}}; w)} 
\end{eqnarray}
Here $b_{-1}$ and $B_{-1}$ are the coefficients of $z^{-1}$ and $\zeta^{-1}$,
resp., in the Laurent expansions of $\psi (z)$ and $\Psi (\zeta)$ in
($\ref{eq.review2D.26}$) and ($\ref{eq.review2D.16}$),
resp., and $\alpha_1$ is the coefficient of the leading order $\zeta$ term in the
conformal mapping $z = w(\zeta) $ in ($\ref{eq.review2D.13}$). Further,  $ T(V;
w) $ is the torque resulting from the  current shape $z = w(\zeta)$
and deformation $V(\sigma)$ of the swimmer, while $ T(V^{\textrm{rot}}; w) $ is the torque
resulted from a rigid rotation of a swimmer also with shape $z = w(\zeta)$ but a
uniform rotational velocity field $V^{\textrm{rot}} (\sigma ; t) = i w
(\sigma;t)$.  More detailed analyses of $U$ and
$\omega$ are given  appendix~$\ref{Appendix:A_tr}$ and
appendix~$\ref{Appendix:A_rot}$.

\item \textit{Force distribution.}
From Table~$\ref{tab.1}$, we see that $f ds = - 2 i \mu d (2 \phi - V)$ on $\partial\Omega$.
The pull-back of the force distribution is:
\begin{eqnarray}
\label{eq.finiteW.9}
 f (\sigma) = - 2 i \mu \big(2 \Phi' (\sigma) - V' (\sigma) \big) \dfrac{d \sigma}{d s} 
= 2 \mu \dfrac{\sigma \big( 2 \Phi' (\sigma) - V' (\sigma) \big)}{\big| w' (\sigma) \big|}
\end{eqnarray}

\item \textit{Power expenditure.}
The power expenditure is calculated by integrating the stress times the velocity on the surface of the
swimmer:
\begin{eqnarray*}
 \mathcal{P} = - \Re \oint_{\partial\Omega} \overline{u} f d s =
 -2 \mu \Im \int_{S^1} \overline{V (\sigma)} \big( 2 \Phi' (\sigma) - V' (\sigma) \big) d \sigma
\end{eqnarray*}
With boundary condition generally given as $ V (\sigma) = \sum_{n \neq 0} \lambda_n \sigma^{ n} $,
together with the expansion of $\Phi (\sigma)$ as given in  ($\ref{eq.review2D.15}$),
we have
\begin{eqnarray}
\label{eq.finiteW.10}
  \mathcal{P} = 4 \pi \mu \sum_{n \geq 1} n \Big( \big| \lambda_{ n} \big|^2 + 2 A_{-n} \overline{\lambda_{-n} } - \big| \lambda_{-n} \big|^2 \Big)
\end{eqnarray}
We define the average power expenditure within 
a period to be
\begin{eqnarray}
\label{eq.finiteW.defmeanpower}
 \widetilde{\mathcal{P}} = \dfrac{1}{T} \int_0^T \mathcal{P} (t) d t  
\end{eqnarray}
%
%
%

\item \textit{Performance.}
We define the performance of the swimmer as 
\begin{eqnarray}
\label{eq.finiteW.defE}
 E = \dfrac{\widetilde{U}}{\widetilde{\mathcal{P}}}= \dfrac{Tr(T)}{ \int_0^T \mathcal{P} (t) d t }
\end{eqnarray}
$E$ measures the distance traveled in one period divided by power expended in a period.

\item \textit{Area of the swimmer.}  
We usually require that the total mass of
the swimmer be constant for cells swimming through a fluid by shape changes. For
a swimmer of constant density in $2$D this becomes an area conservation
constraint.  Suppose that the shape changes of the swimmer are given by
 ($\ref{eq.review2D.13}$), then the swimmer's area  is
\begin{eqnarray*}
  \textrm{Area} (t) = \dfrac{1}{2} \Im \oint \overline{w} d w. 
\end{eqnarray*}
By a direct calculation we obtain
\begin{eqnarray}
\label{eq.finiteW.11}
 \textrm{Area} (t) 
= \pi \big( \big| \alpha_1 \big|^2 - \big| \alpha_{-1} \big|^2 - 2 \big| \alpha_{-2} \big|^2 - \cdots - n \big| \alpha_{-n} \big|^2 - \cdots \big)
\end{eqnarray}
For incompressible swimmers, we require that $\textrm{Area} (t) \equiv \textrm{Constant}$.

\end{enumerate}

A list of scales and units of these physical quantities is given in Table~\ref{tab.review2D.1}.

\begin{table}[htbp]
\renewcommand{\arraystretch}{1.25}
\caption{A list of physical quantities}
\vspace{0.2cm}
\centerline{
\begin{tabular}{|l|l|l|l|}
\hline
                  & Notation                 & Scale                  & Unit\\
\hline
Length            &                          & $R$                    & $\mu m$ \\
\hline
Area              &                          & $ R^2$                  & $\mu m^2$ \\
\hline
Time              & $t$                      & $T$                    & $s$   \\
\hline
Mean velocity     & $\widetilde{U}$           & $R/T$                  & $\mu m / s$  \\
\hline
Force density     & $f$                      & $\mu/T$                & $pN / \mu m$  \\
\hline
Mean power        & $\widetilde{\mathcal{P}}$ & $\mu R^2 / T^2$        & $pN \cdot \mu m / s$  \\
\hline
Performance        & $E$                      & $T/ (\mu R)$               & $1 / p N $  \\
\hline
\end{tabular}
}

\label{tab.review2D.1}
\end{table}

\section{Shapes for conformal mappings with finitely many terms}
\label{Sec.1.5}

In general the Fredholm integral  ($\ref{eq.review2D.22}$) cannot be
solved analytically for an arbitrary conformal mapping $w (\zeta)$. However when
$w (\zeta)$ has only finitely many terms, we find that
 ($\ref{eq.review2D.22}$) can be analytically simplified to a linear
relation between the coefficients of $\Phi$ and $V^+$. In view of the fact that
any conformal mapping can be approximated by truncating its Laurent expansion,
this approach is sufficient in the study of $2$D Stokes flow swimming problems.

\begin{figure}[htbp]
\centering
\includegraphics[width=0.9\textwidth]{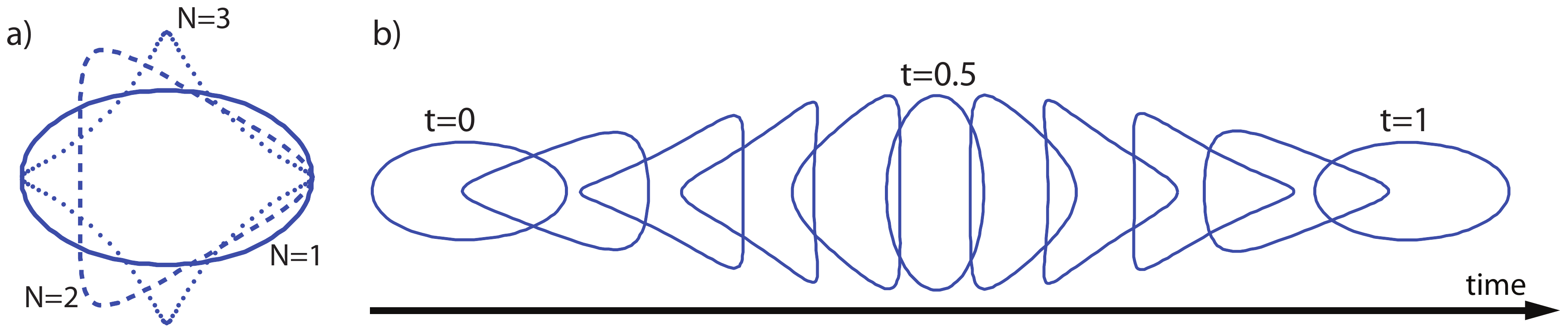}
\caption{The relationship between the conformal mapping and the shape of the swimmer:
a): Shapes determined by $w(\zeta) = 4 \zeta + \zeta^{-N}$, where $N = 1$ (solid line, ellipse),
$N=2$ (dashed line, triangle), or $N =3$ (dotted line, quadrilateral). 
b): A sequence of shape changes defined by  $w (\zeta;t) = 3 \zeta + \cos(2 \pi t) \zeta^{-1} - \sin (2 \pi t)  \zeta^{-2}$.
}
\label{fig.PicConformalShape}
\end{figure}

We consider a sequence of shape changes whose corresponding conformal mappings always 
have Laurent expansions up to $(-N)$-th order. Then 
\begin{eqnarray}
\label{eq.finiteW.1}
  w (\zeta;t) = \alpha_1 (t) \zeta + \alpha_0 (t) + \dfrac{\alpha_{-1} (t)}{\zeta} + \dfrac{\alpha_{-2} (t)}{\zeta^2} + \cdots + \dfrac{\alpha_{-N} (t)}{\zeta^N}
\end{eqnarray}
In general, the $\zeta^{-n}$ term with $n>0$ gives $n+1$
angles along the periphery of the swimmer. Figure~$\ref{fig.PicConformalShape}$ gives an illustration of the relation
between the conformal mapping and the shape of the swimmer.

We assume that the boundary condition is in the general form of 
$V (\sigma;t) = \sum \lambda_n (t) \sigma^n$, and we first prove the following
result in some particular cases.
\begin{Lemma}
\label{Prop.finiteW.1}
Suppose that the conformal mapping $w (\zeta)$ is given by
 ($\ref{eq.finiteW.1}$) and the boundary condition is given as $V
(\sigma) = \sum \lambda_n \sigma^n$.  Let $-V^+ (\zeta)$ be the analytic part of
$V (\zeta)$ when $|\zeta| > 1$, which has the form  $ - V^+ (\sigma; t) = \sum_{n \leq 0}
\lambda_n (t) \sigma^n$.  If there exists a function $f_N (\zeta)$ analytic on
$|\zeta| < 1$, continuous on $|\zeta| \leq 1$, and
\begin{eqnarray*}
 f_N (\sigma) \Big|_{\sigma \in S^1} = \sigma^N \dfrac{w (\sigma)}{\overline{w' (\sigma)}} 
\end{eqnarray*}
for some $N \in \mathbb{Z}^+$, $N \geq 2$, then for any $- V^+ (\zeta)$ with Laurent series such that
$\lambda_{-1} = \lambda_{-2} = \cdots = \lambda_{ - (N - 2)} = 0$  for $N > 2$ (if
$N = 2$ there are no restrictions on the coefficients),
the solution to the boundary value problem  ($\ref{eq.review2D.18}$) is given by
\begin{eqnarray*} 
 \Phi (\zeta) &=&  - V^+ (\zeta)    \\  
 \Psi (\zeta) &=&  \dfrac{\overline{w } (1/\zeta)}{w' (\zeta)} V^{+'} (\zeta) - \overline{V^- } (1/\zeta)
\end{eqnarray*}
for $|\zeta| \geq 1$.
\end{Lemma}
Lemma~$\ref{Prop.finiteW.1}$ can be proved by direct calculation as follows. 
\begin{proof}
 The function  $- V^+$ has the representation
\begin{eqnarray*}
 - V^+ (\zeta) = \dfrac{\lambda_{- (N - 1)}}{\zeta^{N - 1}} + \dfrac{\lambda_{-N}}{\zeta^N} + \cdots + \dfrac{\lambda_{-n}}{\zeta^n} + \cdots
\end{eqnarray*}
Suppose that $\Phi (\zeta) = - V^+ (\zeta)$. Then on the unit circle $S^1$, we have
\begin{eqnarray*}
 \overline{\Phi' (\sigma)} = - \sigma^N \big[ \big( N - 1 \big) \overline{\lambda_{ - (N - 1)}} + N  \overline{\lambda_{-N}} \sigma + \cdots + n  \overline{\lambda_{-n}} 
\sigma^{n - N + 1} + \cdots  \big]
\end{eqnarray*}
and the integral term in  ($\ref{eq.review2D.22}$) becomes 
\begin{eqnarray*}
& & \dfrac{1}{2 \pi i} \int_{S^1} \dfrac{w (\sigma)}{\overline{w' (\sigma)}} \dfrac{\overline{\Phi' (\sigma)}}{\sigma - \zeta} d \sigma  \\
&=& - \dfrac{1}{2 \pi i}\int_{S^1} \dfrac{f_N (\sigma)}{\sigma - \zeta} 
\big[ \big( N - 1 \big) \overline{\lambda_{- (N - 1)}} + N  \overline{\lambda_{-N}} \sigma + \cdots + n  \overline{\lambda_{-n}} 
\sigma^{n - N + 1} + \cdots  \big] d \sigma.
\end{eqnarray*}
Since the function
$$ g(\sigma) = f_N (\sigma) [ ( N - 1 ) \overline{\lambda_{- (N - 1)}} + N
\overline{\lambda_{-N}} \sigma + \cdots + n \overline{\lambda_{-n}} \sigma^{n - N
+ 1} + \cdots ]
$$
 is analytic on $| \sigma | < 1$ and  continuous on $| \sigma | \leq 1$, and
 since  $| \zeta | >1$, the integral 
$$
\dfrac{1}{2 \pi i}  \int_{S^1} \dfrac{g (\sigma)}{\sigma - \zeta} d \sigma = 0
$$
by the Cauchy Integral Theorem. Hence the left hand
side of  ($\ref{eq.review2D.22}$) reduces to $\Phi (\zeta) = - V^+
(\zeta)$.  By substituting this result into  ($\ref{eq.review2D.18}$),  we
obtain the expression of $\Psi$ as in the lemma. Q.E.D.

\end{proof}

With $w$ given by  ($\ref{eq.finiteW.1}$) we have
\begin{eqnarray*}
 \myfrac[3pt]{w (\sigma)}{\overline{w' (\sigma)}} =
\myfrac[3pt]{\alpha_1 \sigma + \alpha_0 + \alpha_{-1} \sigma^{-1} + \alpha_{-2} \sigma^{-2} + \cdots + \alpha_{-N} \sigma^{-N}
}{\overline{\alpha_1} - \overline{\alpha_{-1}} \sigma^2 - 2 \overline{\alpha_{-2}} \sigma^3 - \cdots - N \overline{\alpha_{-N}} \sigma^{N+1} }
\end{eqnarray*}
hence we may take 
\begin{eqnarray} 
\nonumber
 f_N (\zeta) &=& \zeta^N 
\myfrac[3pt]{\alpha_1 \zeta + \alpha_0 + \alpha_{-1} \zeta^{-1} + \alpha_{-2} \zeta^{-2} + \cdots + \alpha_{-N} \zeta^{-N}
}{\vspace*{3pt}\overline{\alpha_1} - \overline{\alpha_{-1}} \zeta^2 - 2
\overline{\alpha_{-2}} \zeta^3 - N \overline{\alpha_{-N}} \zeta^{N+1} }   \\
\label{eq.finiteW.5} 
&=& \myfrac[3pt]{\alpha_{-N} + \alpha_{- (N - 1)} \zeta + \cdots + \alpha_{-1} \zeta^{N-1} + \alpha_0 \zeta^N + \alpha_1 \zeta^{N + 1}
}{\overline{\alpha_1} - \overline{\alpha_{-1}} \zeta^2 - 2
\overline{\alpha_{-2}} \zeta^3 - \cdots - N \overline{\alpha_{-N}} \zeta^{N+1} }    
\end{eqnarray}
If  no singularity of $f_N$ lies inside the unit disk $|\zeta| < 1$, then $f_N $ is analytic 
on $|\zeta| < 1$ and we may apply Lemma~$\ref{Prop.finiteW.1}$ to obtain the
solution of the boundary condition constraint ($\ref{eq.review2D.12}$).

The Laurent expansion of any function $\Phi (\zeta)$ that is analytic on
 $|\zeta| > 1$, continuous on $|\zeta| = 1$, has the form of
  ~($\ref{eq.review2D.15}$).  Let $\mathcal{S}$ be the set 
\begin{eqnarray*}
 \mathcal{S} = \Big\{ \Phi (\zeta) = \dfrac{A_{-1}}{\zeta} + \dfrac{A_{-2}}{\zeta^2} + \dfrac{A_{-3}}{\zeta^3} + \cdots ;  \sum_{k \geq 1} k | A_{-k} | < + \infty   \Big\}
\end{eqnarray*}
 of all functions $\Phi (\zeta)$ whose coefficients of the Laurent expansion
satisfy $\sum_{k \geq 1} k | A_{-k} | < + \infty$ and vanish at infinity. $\mathcal{S}$ is
 a Banach space endowed with the norm $ \parallel \Phi \parallel = \sum_{k 
\geq 1} k | A_{-k} | $ \citep{Chambrion:2011:LCS}.

Now with $w$ given by  ($\ref{eq.finiteW.1}$), we define a conjugate linear operator $\bm{K} (w)$ on $\mathcal{S}$ as  
\begin{eqnarray}\label{eq.finiteW.6}
 \big[ \bm{K} (w) \circ  \Phi \big] (\zeta) = \dfrac{1}{2 \pi i} \int_{S^1} \dfrac{w (\sigma)}{\overline{w' (\sigma)}} \dfrac{\overline{\Phi' (\sigma)}}{\sigma - \zeta} d \sigma
\end{eqnarray}
Let $\mathcal{S}_0 = \{ A_0 \in \mathbb{C}\}$, then $\bm{K} (w)$ can be 
extended to $\mathcal{S} \oplus \mathcal{S}_0$ such that $\bm{K} (w) |_{\mathcal{S}_0} \equiv 0$ 
in accordance with  ($\ref{eq.finiteW.6}$).
Now for any $\Phi \in \mathcal{S} \oplus \mathcal{S}_0 $,  ($\ref{eq.review2D.22}$) can be written as
\begin{eqnarray*}
 \big[ \bm{I} + \bm{K} (w) \big] \Phi = - V^+
\end{eqnarray*}
where $\bm{I}$ is the identity operator. 
For $n \in \mathbb{Z}^+$, let 
\begin{eqnarray*}
 \mathcal{S}_n^{\Re} = \big\{ \dfrac{r}{\zeta^n} ; r \in \mathbb{R} \big\} \qquad \textrm{and} \qquad
\mathcal{S}_n^{\Im} = \big\{ \dfrac{i r}{\zeta^n} ; r \in \mathbb{R} \big\} 
\end{eqnarray*}
It is easily seen that
$\mathcal{S}_n^{\Re}$'s and $\mathcal{S}_n^{\Im}$'s are real $1$-dimensional linear subspaces of $\mathcal{S}$.
With $f_N$ given by  ($\ref{eq.finiteW.5}$), by Lemma~$\ref{Prop.finiteW.1}$, 
$\bm{K} (w)$ vanishes on each $\mathcal{S}_n^{\Re}$ 
and $\mathcal{S}_n^{\Im}$ for $n = N-1, N, N+1, \cdots$, and thus each $\mathcal{S}_n^{\Re}$ or $\mathcal{S}_n^{\Im}$ is invariant under 
the operator $\bm{I + K} (w)$, i.e.,
\begin{eqnarray*}
 \big[ \bm{I} + \bm{K} (w) \big] \big|_{\mathcal{S}_n^{\Re}} = \bm{I}_{\mathcal{S}_n^{\Re} } ; \qquad \qquad 
 \big[ \bm{I} + \bm{K} (w) \big] \big|_{\mathcal{S}_n^{\Im}} = \bm{I}_{\mathcal{S}_n^{\Im} }
\end{eqnarray*}

Define
\begin{eqnarray*}
\mathcal{S}_N :=   \textrm{Span}_{\mathbb{R}}\{\zeta^{-n}, i \zeta^{-n}\}_{n= - 1}^{- (N-2)} =
\bigoplus_{n = - 1}^{- (N - 2)} \big( \mathcal{S}_n^{\Re} \bigoplus \mathcal{S}_n^{\Im} \big)
\end{eqnarray*}
which is a real $2 (N - 2)$-dimensional linear subspace of $\mathcal{S}$, and
\begin{eqnarray*}
 \{ \zeta^{-1}, \zeta^{-2}, \cdots, \zeta^{N-2}, i \zeta^{-1}, i \zeta^{-2}, \cdots,  i \zeta^{N-2} \} 
\end{eqnarray*}
consists a basis of $\mathcal{S}_N$. 
If $f_N$ is analytic on $|\zeta| < 1$, then the action of $\bm{K} (w)$ on this
basis can be expressed as
\begin{eqnarray*}
\big[ \bm{K} (w) \big]\circ \zeta^{-n} &=& - \dfrac{n}{2 \pi i}\int_{S^1} \dfrac{f_N (\sigma)}{\sigma - \zeta} \dfrac{1}{\sigma^{N - n -1}} d \sigma  
= \dfrac{n}{\zeta^{N - n - 1}} \sum_{ k = 0}^{N - n - 2} \dfrac{f_N^{(k)} (0)}{k !} \zeta^k \\
\big[ \bm{K} (w) \big] \circ \big( i \zeta^{-n} \big) &=& \dfrac{n}{2 \pi}\int_{S^1} \dfrac{f_N (\sigma)}{\sigma - \zeta} \dfrac{1}{\sigma^{N - n -1}} d \sigma  
= - \dfrac{i n}{\zeta^{N - n - 1}} \sum_{ k = 0}^{N - n - 2} \dfrac{f_N^{(k)} (0)}{k !} \zeta^k
\end{eqnarray*}
for $n = 1, 2, \cdots, N-2 $.
Notice that the sum in the above equations is simply the first $N-n-1$ terms of the Laurent 
expansion of $f_N (\zeta)$ at $\zeta = 0$.
It is easily seen that on $ \mathcal{S}_N $,
the operator $\bm{K} (w)$ has a matrix representation as
\begin{eqnarray*}
\big[ \bm{K} (w) \big] \circ (\zeta^{-1}, \zeta^{-2}, \cdots, \zeta^{- (N - 2)})^T &=& 
K (w) \cdot  (\zeta^{-1}, \zeta^{-2}, \cdots, \zeta^{- (N - 2)})^T  \\
\big[ \bm{K} (w) \big] \circ (i \zeta^{-1}, i \zeta^{-2}, \cdots, i \zeta^{- (N - 2)})^T &=& 
- K (w) \cdot  (i \zeta^{-1}, i \zeta^{-2}, \cdots, i \zeta^{- (N - 2)})^T
\end{eqnarray*}
where $K (w)$ is a $(N - 2) \times (N - 2)$ matrix with the form 
\begin{eqnarray}
\label{eq.finiteW.7}
 K  (w)= \left( \begin{array}{cccccccc}
 \dfrac{f_N^{(N-3)} (0)}{(N - 3)!}   &  \dfrac{f_N^{(N-4)} (0)}{(N - 4)!}    &  \cdots & \cdots & \cdots & \dfrac{f_N'' (0)}{2!}  & \dfrac{f_N' (0)}{1!}  & \dfrac{f_N (0)}{0!} \\
 \dfrac{2 f_N^{(N-4)} (0)}{(N - 4)!} &  \dfrac{2 f_N^{(N-5)} (0)}{(N - 5)!}  &  \cdots & \cdots & \cdots &  \dfrac{2f_N' (0)}{1!} & \dfrac{2 f_N (0)}{0!} & 0            \\
 \dfrac{3 f_N^{(N-5)} (0)}{(N - 5)!} &  \dfrac{3 f_N^{(N-6)} (0)}{(N - 6)!}  &  \cdots & \cdots & \cdots &  \dfrac{3f_N  (0)}{0!} & 0                     & 0            \\
             \vdots      &   \vdots       &          &         & \cdotp  & 0        & 0    & \vdots       \\
            \vdots       &  \vdots        &          & \cdotp  & \cdotp  & \vdots   &  \vdots   & \vdots      \\
             \vdots      &   \vdots       &  \cdotp  & \cdotp  &         & \vdots   &  \vdots   & \vdots       \\
            \dfrac{(N - 3) f_N' (0)}{1!}   &   \dfrac{(N - 3) f_N (0)}{0!}  &  0       &  \cdots &  \cdots &  0       & 0    & 0            \\
            \dfrac{(N - 2) f_N (0)}{0!}    &   0            &   \cdots & \cdots  &  \cdots & 0        &  0   & 0            \\
            \end{array}  \right)
\end{eqnarray}

This can be summarized as follows.
\begin{Lemma}

\label{Prop.finiteW.2}
Suppose that the conformal mapping $w (\zeta)$ is given by  ($\ref{eq.finiteW.1}$) and the boundary condition
is given as $V (\sigma) = \sum \lambda_n \sigma^n$.
If there exists a function $f_N (\zeta)$ analytic on $|\zeta| < 1$, continuous on $|\zeta| \leq 1$, and 
\begin{eqnarray*}
 f_N (\sigma) \Big|_{\sigma \in S^1} = \sigma^N \dfrac{w (\sigma)}{\overline{w' (\sigma)}} 
\end{eqnarray*}
for some $N \in \mathbb{Z}^+$, $N \geq 2$, then the integral  ($\ref{eq.review2D.22}$) reduces to
linear relations between coefficients of $\Phi$ and $V$:
\begin{enumerate}
 \item For $n = 0$ or $n \geq N - 1$, $A_{- n } = \lambda_{-n}$;
 \item For $1 \leq n \leq N-2$:
\begin{eqnarray*}
 \big( I_{N - 2} + K (w) \big)\big( \Re A_{-1}, \cdots, \Re A_{- (N - 2)} \big)^T &=& \big( \Re \lambda_{-1}, \cdots, \Re \lambda_{N-2} \big)^T \\
\big( I_{N - 2} - K   (w) \big)\big( \Im A_{-1}, \cdots, \Im  A_{- (N - 2)} \big)^T &=& \big( \Im  \lambda_{-1}, \cdots, \Im  \lambda_{N-2} \big)^T 
\end{eqnarray*}
where $I_{N-2}$ is the $(N-2)$-identity matrix, and $K  (w)$ is the $(N - 2) \times (N - 2)$ matrix 
given by  ($\ref{eq.finiteW.7}$).
\end{enumerate}
\end{Lemma}

Lemma~$\ref{Prop.finiteW.2}$ provides an algorithm for solving the Stokes
equation of an infinite $2$D Stokes flow when the shape deformations have
finitely many terms in the conformal mapping $w (\zeta)$. Once we solve for $ \Phi$
according to Lemma~$\ref{Prop.finiteW.2}$, $\Psi$ is given by the following expression.
\begin{eqnarray}\nonumber
  \Psi (\zeta; t) &=& - \dfrac{\overline{\dot{\alpha}_1}}{\zeta} + \overline{ \big( A_{-1} - \dot{\alpha}_{-1} \big)} \zeta + \overline{\big( A_{-2} - \dot{\alpha}_{-2} \big) }\zeta^2
+ \cdots +\overline{ \big( A_{-N} - \dot{\alpha}_{-N} \big)} \zeta^N  \\ 
\label{eq.finiteW.8}
& & \myfrac[6pt]{\dfrac{\overline{\alpha_1}}{\zeta} + \overline{\alpha_{-1}}
\zeta +\overline{ \alpha_{-2}} \zeta^2 + \cdots +\overline{ \alpha_{-N}}
\zeta^N}{ \alpha_1 
- \dfrac{\alpha_{-1}}{\zeta^2} - \dfrac{2 \alpha_{-2}}{\zeta^3} - \cdots -
\dfrac{N \alpha_{-N}}{\zeta^{N+1}}}
\Big( \dfrac{A_{-1}}{\zeta^2} + \dfrac{2 A_{-2}}{\zeta^3} + \cdots + \dfrac{N A_{-N}}{\zeta^{N+1}} \Big) \quad
\end{eqnarray}

\section{Modeling of swimming Dictyostelium amoebae}
\label{Sec.2.1}

\begin{figure}[htbp]
\centering
\includegraphics[width=1\textwidth]{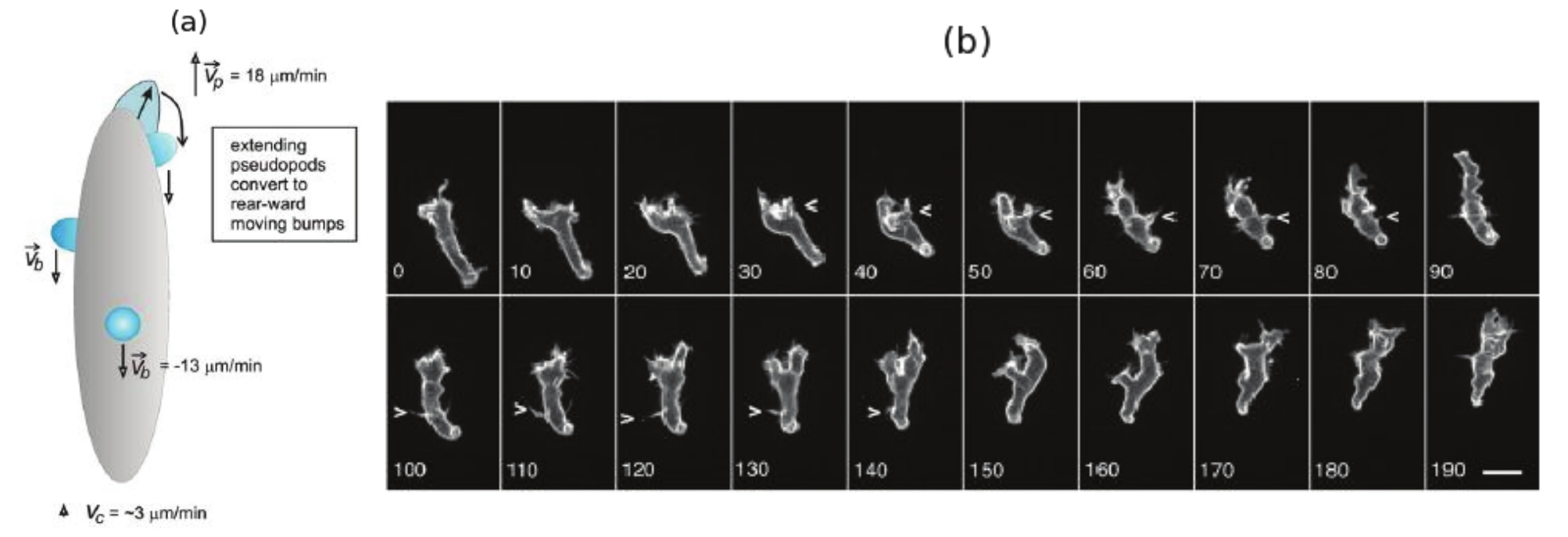}
\caption[Amoebae swim by protrusions]{Amoebae swim by protrusions:
(a) a swimming cell with $3$ protrusions \citep{VanHaastert:2011:ACU};
(b) the shape of an amoeboid as it swims \citep{Barry:2010:DAN}.}
\label{fig.sc.3}
\end{figure}

As mentioned in the Introduction, Dictyostelium amoebae can move in a fluid
environment by a combination of blebbing and protrusions, either of which
involve rapid shape changes and neither of which require attachment to a
substrate.  Experimental observations on the movement of Dd cells reported in
\cite{Barry:2010:DAN,Bae:2010:SDA} and \cite{VanHaastert:2011:ACU} have recorded cell shape
changes, speeds, and periods of the cyclic motion, and we use their data here to
compare with theoretical predictions. The movement usually involves protrusions
that are initiated at the leading edge of the cell and which propagate toward
the rear. Typically the cell body is elongated, and multiple protrusions
propagate along the cell.  van Haastert \citep{VanHaastert:2011:ACU}
reported an average of three protrusions, as illustrated by the cartoon model in
Figure~$\ref{fig.sc.3}$ (a), while from the experimental images
(Figure~$\ref{fig.sc.3}$ (b)) of a swimming Dictyostelium reported in Barry
et.al. \citep{Barry:2010:DAN}, we see that one protrusion travels along one side
of the cell and disappears at the rear of the cell, then another protrusion
appears on the other side and repeats the  process.

Van Haastert \citep{VanHaastert:2011:ACU} observed that the protrusions travel
directly down the cell body and not in a helical fashion. Thus there is no clear
evidence that the cell is rotating around its symmetry axis, and as a result we
consider the 2D model developed in the preceding sections to be a reasonable
simplification of a 3D swimming cell.  In Section~$\ref{Sec.CellShape}$ the
numerical scheme used to construct the shape of the cell is developed.  As we
see from Figure~$\ref{fig.sc.3}$, swimming by extending protrusions is mostly
asymmetric in that they alternate sides, and thus the motion is not
rotation-free and the trajectory of a swimming cell is snake-like rather than
along a straight line. However we begin in Section~\ref{Ddsim} with a simplified
symmetric amoeba, where a pair of side protrusions move down the cell body
symmetrically so as to minimize the mechanical effects resulted from rotation or
cell body twisting. In Section~\ref{Height} we investigate how different cell
and protrusion shapes affect the swimming ability of such a translational
swimmer.  Finally, in Section~\ref{asymm} we consider an asymmetric swimmer
similar to that shown in Figure~$\ref{fig.sc.3}$(b) and compare such a
snake-like swimming style to the symmetric swimming style.

\subsection{Construction of the cell shape}
\label{Sec.CellShape}

To apply the Muskhelishvili method to a swimming cell we first have to obtain
the conformal mapping $w$ corresponding to the cell shape, and then truncate its
Laurent's expansion, leaving only $N$ negative order terms for some $N$. In
general it is difficult to find the conformal mapping analytically that
corresponds to a general shape, yet that of an $n$-polygon can be found by use
of the Schwarz-Christoffel formula \citep{Ahlfors:1978:CA,driscoll2002schwarz}.

\begin{figure}[htbp]
\centering
\includegraphics[width=0.5\textwidth]{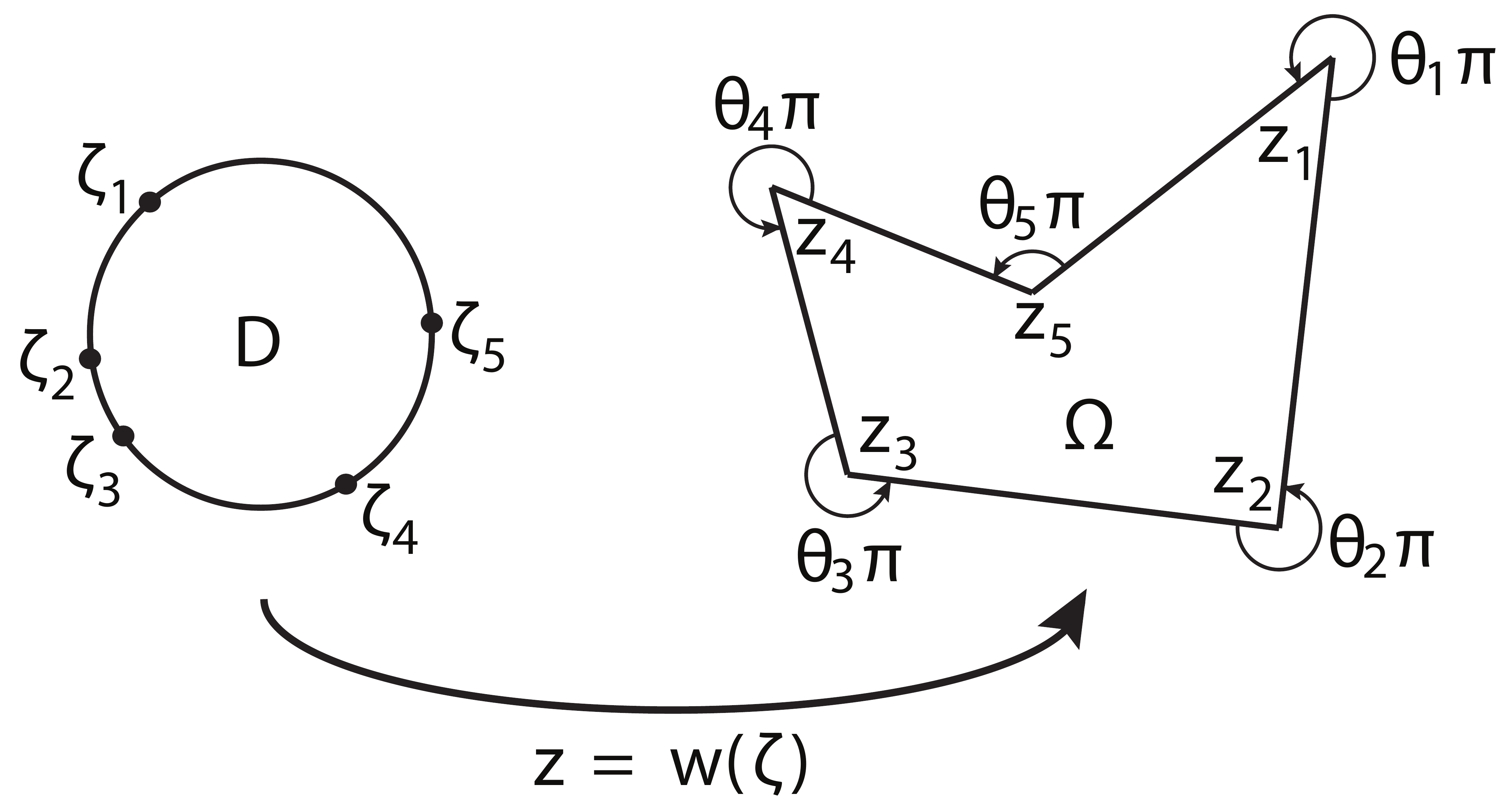}
\caption[Conformal mapping for the exterior region of a polygon]{The conformal mapping for the exterior region of a polygon 
from that of the unit disk.}
\label{fig.sc.1}
\end{figure}

Suppose that we have an $n$-polygon in the $z$-plane with vertices $z_1,
\cdots,z_n$, and the corresponding exterior angles are $\theta_1 \pi,
\cdots,\theta_n \pi$ (Figure~$\ref{fig.sc.1}$). Let $\Omega$ be the interior region
bounded by the polygon.  The conformal mapping from the exterior of the unit
disk $D$ in the $\zeta$-plane to $\Omega^c$ is given by the following
Schwarz-Christoffel (SC) formula \citep{Ahlfors:1978:CA,driscoll2002schwarz}:
\begin{eqnarray}\label{sc.1} z = w (\zeta) = A + C \int^{\zeta} \dfrac{1}{\xi^2}
\prod_{k = 1}^n \big( \xi - \zeta_k \big)^{\theta_k - 1} d \xi \end{eqnarray}
where $z_k = w (\zeta_k)$ and we call $\zeta_k$ the \textit{prevertex} to the
vertex $z_k$ under the conformal mapping $z = w(\zeta)$. Since we are
considering transformation from the exterior region of $D$ to the exterior
region of $\Omega$, the angles should satisfy $ \sum_{k = 1}^n \theta_k = n +
2$.
The prevertices $\zeta_k$ can be numerically approached by using the SC toolbox
given at {\em  http://www.math.udel.edu/\~{}driscoll/SC/}.
Once we have obtained the Schwarz-Christoffel transformation $w_{\textrm{SC}}
(\zeta)$ for a given polygon, we truncate 
its Laurent expansion leaving only $N$ negative order terms, which then has  the form
\begin{eqnarray}
\label{eq.SCN}
w_{\textrm{SCN}} (\zeta) = \alpha_1 \zeta + \alpha_0 + \dfrac{\alpha_{-1}}{\zeta} + \cdots + \dfrac{\alpha_{-N}}{\zeta^N}.
\end{eqnarray}
The image of the unit circle $S^1$ under the truncated conformal mapping
$w_{\textrm{SCN}} $ is a contour approximating the original polygon.  

\begin{figure}[htbp]
\centering
\includegraphics[width=.9\textwidth]{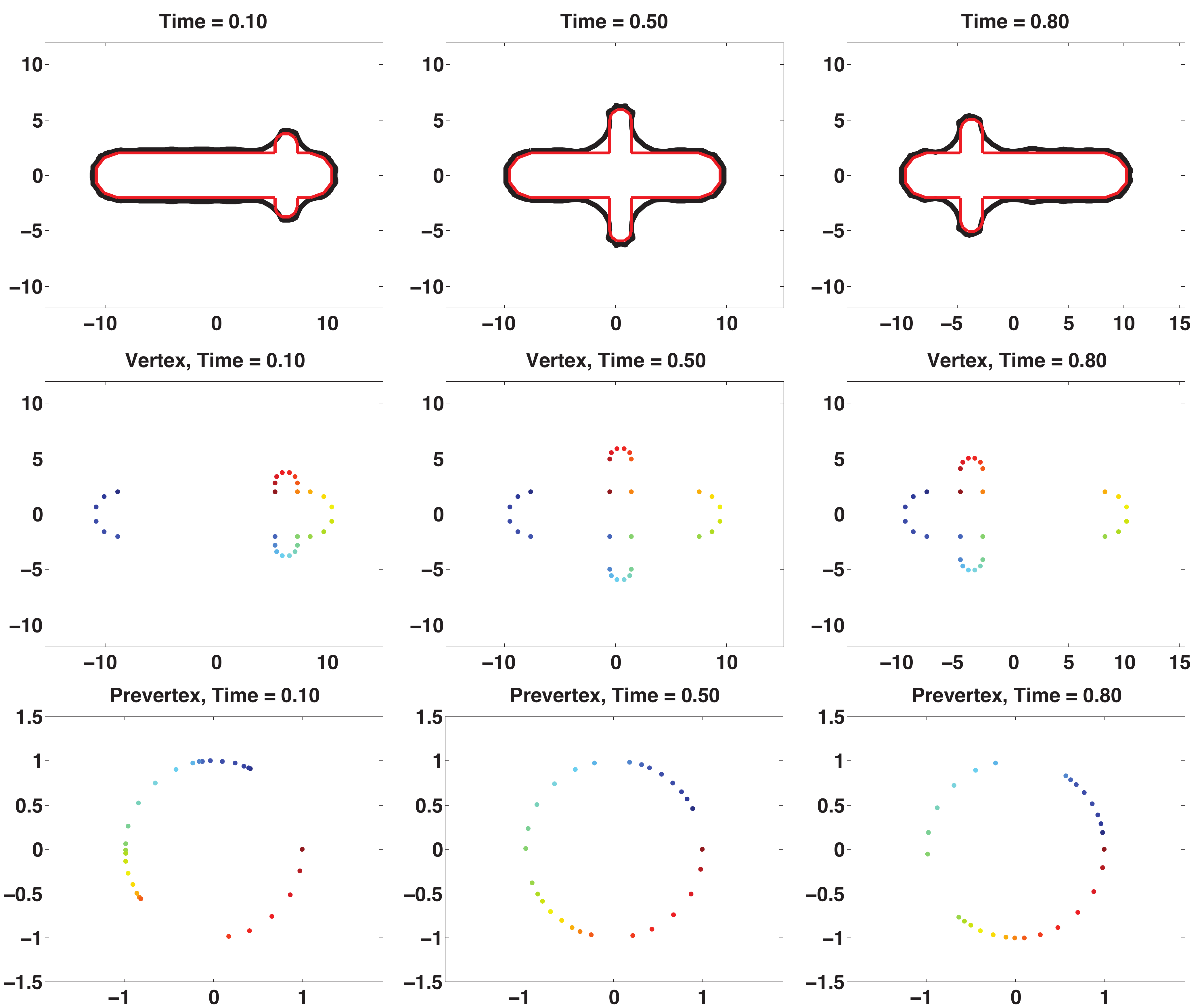}
\caption[Conformal mapping for the exterior region of a polygon]{The conformal mapping for the exterior region of a polygon 
from that of the unit disk.  
The axes are in units of $\mu m$.}
\label{fig.IllCellShape}
\end{figure}

We approach the  construction of the shape of a swimming amoeba as
follows. Instead of using a polygonal discretization  with many nodes on the
boundary, we first construct an "inner skeleton" of the cell with only a few
nodes (Figure~$\ref{fig.IllCellShape}$, red contours), then we obtain its
Schwarz-Christoffel transformation and truncate it  to obtain
$w_{\textrm{SCN}} $. Finally we smooth the image of the unit circle under
$w_{\textrm{SCN}} $ by multiplying the amplitude of  $\alpha_1 $ in
 ($\ref{eq.SCN}$) by a factor $C \in \mathbb{R} $,  $C > 1$, to give
\begin{eqnarray}
\label{eq.SCN1}
w_N (\zeta) = C \alpha_1 \zeta + \alpha_0 + \dfrac{\alpha_{-1}}{\zeta} + \cdots + \dfrac{\alpha_{-N}}{\zeta^N}.
\end{eqnarray}
 This will give us a smoothed contour enclosing the "inner skeleton", and we use
it as the shape of cell (Figure~$\ref{fig.IllCellShape}$, black
contours). Figure~$\ref{fig.IllCellShape}$ illustrates this process, and the
detailed steps of the algorithm are given in the appendix~$\ref{Appendix:Pseudo_Code}$. The panels
in the top row give three snapshots within one cycle. The red contours give the
"inner skeletons", each one is a prescribed polygon, with each of the four
semicircle ends having five nodes. The black contours that are taken as the
current cell shapes, are determined by a conformal mapping of the form of
 ($\ref{eq.SCN}$) with $N=30$ and $C = 1.05$. The panels in the middle
and bottom rows show the distribution of vertices of the polygon ($z_i$ in
Figure~$\ref{fig.sc.1}$) and the distribution of prevertices along the unit
circle ($\zeta_i$ in Figure~$\ref{fig.sc.1}$), with the correspondence relation
given by the color of the dots. From Figure~$\ref{fig.IllCellShape}$ we see that
when the two side protrusions are close to either end of the cell body, some
prevertices are crowded and when the side protrusions are near the middle of the
cell body, the prevertices are more scattered.

In the simulations described later  we do not require strict area conservation
-- instead we require that the area changes be  restricted within a small
range. We define the ratio of area change within one period to be
\begin{eqnarray*}
 \textrm{Ratio of area change} = \dfrac{\textrm{Maximum of area} -
 \textrm{Minimum of area}}{\textrm{Average of area}} 
\end{eqnarray*}
and we require that the ratio be  $\leq 0.1$.

\subsection{Simulation results of swimming Dictyostelium amoebae}
\label{Ddsim}

We use the data for swimming amoebae from
\citep{VanHaastert:2011:ACU,Barry:2010:DAN}.  Though they are both Dictyostelium
amoebae, they have different sizes. For simplicity we will refer to them as
``van Haastert's cell'' (Figure~$\ref{fig.sc.3}$ (a)) and ``Barry's cell''
(Figure~$\ref{fig.sc.3}$ (b)) hereafter.  Their data is shown in
Table~$\ref{tab.sc.1}$.  

\begin{table}[htbp]
\caption{Experimental data for Van Haastert's cell \protect\citep{VanHaastert:2011:ACU} and Barry's cell \protect\citep{Barry:2010:DAN}.}
\vspace{0.2cm}
\renewcommand{\arraystretch}{1.5}
\centerline{
\begin{tabular}{|l|l|l|}
\hline
& Van Haastert & Barry \\
\hline
Maximum cell body length & $\sim 25 \mu m$ & $\sim 22 \mu m$\\
\hline
Average cell body width & $\sim 6 \mu m$  & $\sim 4 \mu m$  \\
\hline
Maximum protrusion height & $\sim 2 \mu m$ & $\sim 4 \mu m$ \\
\hline
Average protrusion width & $\sim 2 \mu m $ & $\sim 2 \mu m $\\
\hline
Period of a stroke & $\sim 1 \textrm{min}$ & $\sim 1.5 \textrm{min}$\\
\hline
\end{tabular}
}
\label{tab.sc.1}
\end{table}

\begin{figure}[htbp]
\centering
\includegraphics[width=.95\textwidth]{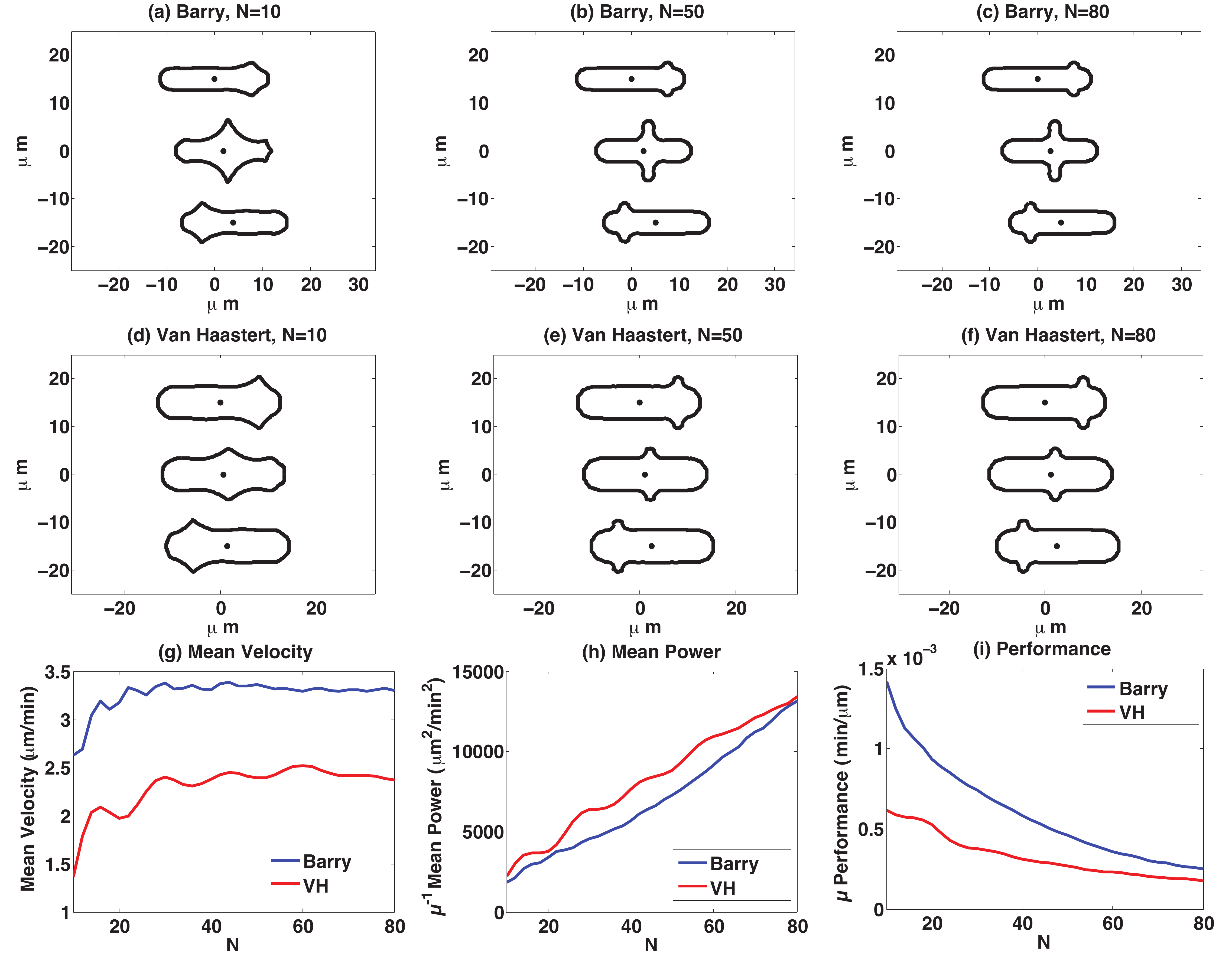}
\caption[Swimming and power of van Haastert's cell]{(a-c) The shapes of Barry's
  cell within a cycle for different values of $N$.  
  (d-f) The shapes of van Haastert's cell within a cycle
  for different values of $N$. 
  In (a-f), the top rows are the snapshots at the beginning of a cycle,
  the middle rows are at half of the cycle, and the bottom rows are at the end
  of the cycle. For the periods of the two cells, we have $T_B = 1.5 \
  \textrm{min}$ for Barry's cell and $T_H = 1 \ \textrm{min}$ for van Haastert's
  cell.  (g-i) A comparison of the mean velocity $\widetilde{U}$, mean power
  $\widetilde{\mathcal{P}}$ and performance $E$ of the two cells for $N
  \in[10,80]$. }
\label{fig.BarryHaastert_N}
\end{figure}

We use the numerical methods discussed in Section~$\ref{Sec.CellShape}$ to
generate sequences of shapes based on data given in Table~$\ref{tab.sc.1}$.
The numerical results are presented in Figure~$\ref{fig.BarryHaastert_N}$.
The shapes of  Barry's cell within a cycle for different values of $N$ are shown in 
Figure~$\ref{fig.BarryHaastert_N}$(a-c), while those of van Haastert's cell are shown in 
Figure~$\ref{fig.BarryHaastert_N}$(d-f). Figure~$\ref{fig.BarryHaastert_N}$(g-i) compare the
mean velocity $\widetilde{U}$, mean power $\overline{\mathcal{P}}$ and performance $E$
of the two cells for $N \in[10,80]$. 

A number of conclusions can be drawn from these simulations, as listed below. 

\begin{enumerate}
 \item \textit{Protrusion shape.} First we consider how the number of terms $N$
in the conformal mappings $w_N$ affects the shapes of the swimmers. As we
mentioned earlier, in general the $\zeta^{-n}$ term gives $n$ angles along the
periphery of the cell. From Figure~$\ref{fig.BarryHaastert_N}$(a - f) we see
that swimmers corresponding to $w_N $ with larger $N$ have more rounded heads in
the protrusions, while those corresponding to $w_N $ with smaller $N$ tend to
have sharper heads; another important difference in the shapes of the
protrusions is that the connecting parts between the cell body and the
protrusions are smoother for smaller $N$ while more abrupt for larger $N$.
Figure~$\ref{fig.sc.4}$ gives an enlarged view of the protrusion regions of both
cells with different $N$.

\begin{figure}[htbp]
\centering
\includegraphics[width=.9\textwidth]{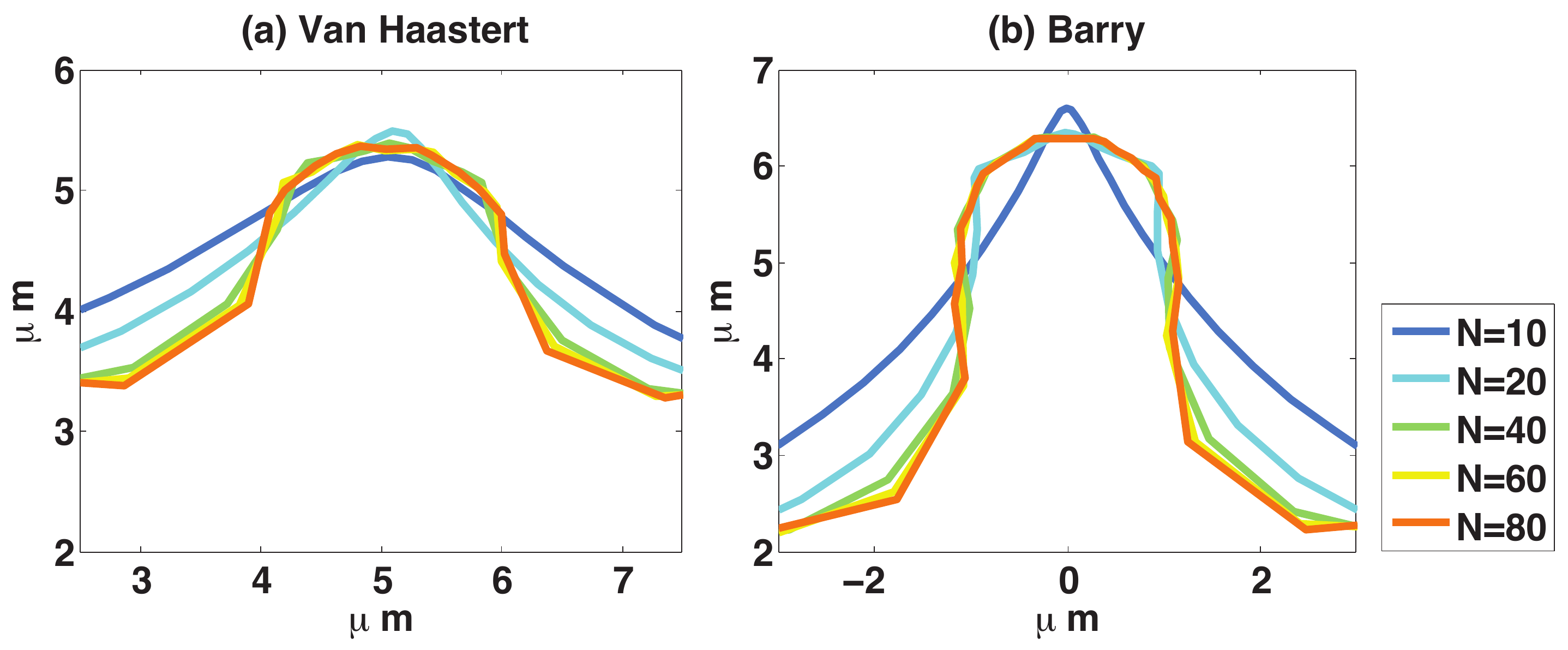}
\caption{The  areas of the protrusions 
in the models of van  Haastert's and Barry's cells for different $N$:
(a)van Haastert's cell; (b) Barry's cell.}
\label{fig.sc.4}
\end{figure}

\item \textit{Velocity.} For either cell, the mean velocity $\widetilde{U}$
increases rapidly with $N$ when $N$ is small, but 
$\widetilde{U}$ does not change much for even larger $N > \sim 40$
(Figure~$\ref{fig.BarryHaastert_N}$(g)). For $N \geq 40$, $\widetilde{U}$ of van
Haastert's cell fluctuates within a range $2.37 - 2.52 \mu m / \textrm{min}$
while $\sim 3.30 \mu m / \textrm{min}$ for Barry's cell. Taking into account of
our observation of the relation between $N$ and protrusion shapes, our
simulation results indicate that \textit{abrupt protrusions with rounded heads
may enhance the swimming speed}.

\item \textit{Power.} Unlike the mean velocity $\widetilde{U}$ which has a
maximum as $N$ increases, the mean power $\overline{\mathcal{P}}$ continues
increasing as $N$ increases (Figure~$\ref{fig.BarryHaastert_N}$(h)). In
particular, if we observe Figures~$\ref{fig.BarryHaastert_N}$(b,c,e,f), we see
that the shapes of the same cell for $N=50$ and $N=80$ are quite similar, yet with
more terms in $w_N$ it requires much more power.

\item \textit{Performance.} Figure~$\ref{fig.BarryHaastert_N}$(i) clearly shows
that performance decreases as more terms $N$ in $w_N$ are
involved. Incorporating the observation of bump shapes, we find that
\textit{smoother protrusions tend to lead to swimming with better performance
though it might be slower}.

\item \textit{Comparing with experimental data.} van Haastert
\citep{VanHaastert:2011:ACU} reported that the swimming velocity of a typical
cell is $\sim~ 3 \mu m / \textrm{min}$, while our model predicts $ \sim 1.36 -
2.52 ~\mu m / \textrm{min}$.  Barry at.el. \citep{Barry:2010:DAN} reported that
the linear speed of the cells has a range of $2 - 8.4~\mu m/ \textrm{min}$ with
an average of about $4.2 \mu m/ \textrm{min}$, comparing to a range of $\sim
2.63 - 3.31 \mu m/ \textrm{min}$ as given by our numerical simulations. So far
there are no experimental data regarding the power or performance of the
swimming amoebae, and based on the data of swimming speed collected from
experiments, we believe our model is reasonable.

\item \textit{How cell shapes affect the swimming behavior.} Both cells are the
same kind of Dictyostelium amoebae, but from the above results and observations
we clearly see that different cell shapes and sizes lead to the difference in
their swimming behavior:  as compared with  van Haastert's cell, Barry's cell is more
slender, with higher protrusions, and smaller in size, and the  simulation results
show that the average area for van Haastert's cell ranges within $170 \mu m^2 -
174 \mu m^2$, depending on the value of $N$, while the average area for Barry's
cell is only within $103 \mu m^2 - 106 \mu m^2$.
Figure~$\ref{fig.BarryHaastert_N}$ indicates that Barry's cell exhibits faster
swimming and better performance. This inspires us to study how the sizes of the
cell body and the protrusions affect the swimming behavior of the cell, which
will be discussed in detail in the following section.

\end{enumerate}

\subsection{Effects of the protrusion height and cell body shapes on swimming}
\label{Height}

\begin{figure}[htbp]
\centering
\includegraphics[width=1\textwidth]{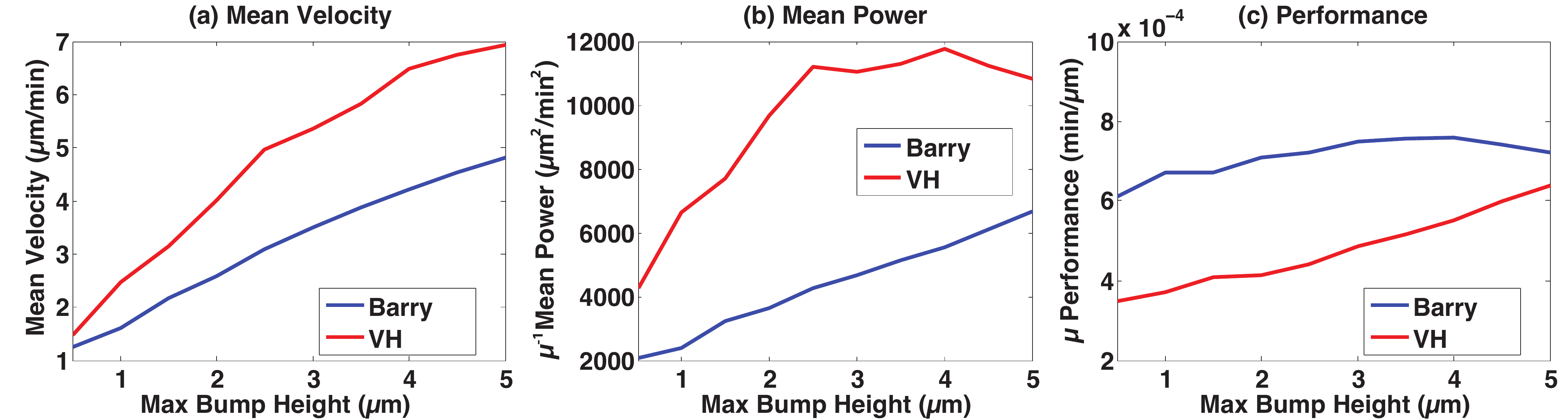}
\caption[Effects of the protrusion height on swimming (I)]{Effects of the protrusion height on swimming  
for van Haastert's and Barry's cell: 
(a): mean velocity $\sim $ maximum protrusion height; 
(b): mean power $\sim$ maximum protrusion height; 
(c): performance $\sim$ maximum protrusion height. }
\label{fig.sc.19}
\end{figure}

We first study the effects of the protrusion shape on the swimming behavior of
the Dictyostelium amoebae.  As above we use the data for the sizes of different
characteristic features of a cell in
\cite{VanHaastert:2011:ACU} and \cite{Barry:2010:DAN}, which are presented in
Table~$\ref{tab.sc.1}$.  As for the conformal mappings, we truncate them at
$N=30$ so as to generate protrusions that are neither too smooth nor too
abrupt. At each time step we adjust the cell body length to compensate for the
area changes caused by the emergence and disappearance of the protrusions, so as
to control the area change of the whole cell within a small range.  We test for
different maximum protrusion heights, ranging from $1.5 \mu m - 5 \mu
m$. Figure~$\ref{fig.sc.19}$ gives the relationship between the mean velocity,
mean power, and performance and the maximum protrusion height for both Barry's
and van Haastert's cells. First we see that the mean velocity increases
significantly as the protrusion becomes taller (Figure~$\ref{fig.sc.19}$
(a)). On the other hand the mean power for Barry's cell increases steadily as
the protrusion height increases, while for van Haastert's cell the mean power
first increases but for taller protrusions it approaches a maximum
(Figure~$\ref{fig.sc.19}$ (b)). Finally for the performance, it turns out that
Barry's cell,which is smaller, always performs better than van Haastert's cell,
though it swims slower than van Haastert's (Figure~$\ref{fig.sc.19}$
(c)). Moreover, while the performance for van Haastert's cell increases as the
protrusion grows higher, the protrusion height does not have much effect on the
performance of Barry's cell.

Next we consider the effects of the cell body shape on the swimming behaviors of
Dictyostelium amoebae by varying the length to width ratio of cells.  We
consider protrusions with the same height ($\sim 3 \mu m$) and width ($\sim 2
\mu m$) and let the shape deformations have the same period ($= 1
\textrm{min}$).  Moreover, to reduce the computational effort we keep the
protrusion height constant through the whole cycle, {\em ie,} we do not consider the
emergence, growth and disappearance processes of the protrusions as in previous
simulations that led to Figure~$\ref{fig.sc.19}$.  To make the comparison fair,
we control the average area of each cell within a certain range ($\sim 176 \mu
m^2 - 205 \mu m^2$).

\begin{figure}[htbp]
\centering
\includegraphics[width=1\textwidth]{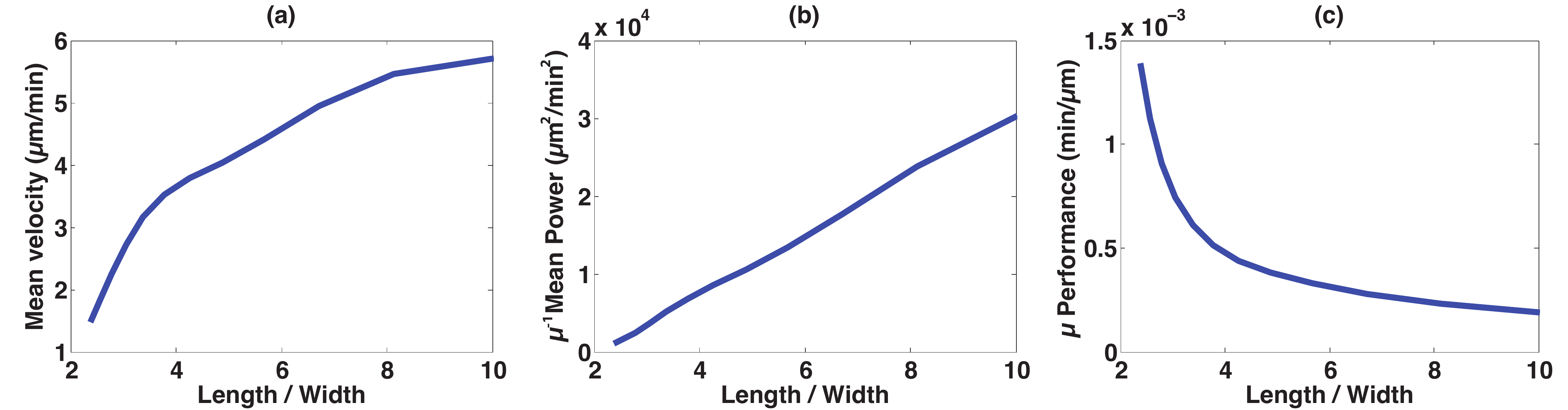}
\caption[Effects of cell body shapes to swimming]{Effects of the cell body shape
  on swimming:
(a): mean velocity $\sim $ $R_s$; 
(b): mean power $\sim$ $R_s$; 
(c): performance $\sim$ $R_s$.  }
\label{fig.sc.21}
\end{figure}

We define the aspect ratio $R_s$ of the cells  as
\begin{eqnarray*}
 R_s = \dfrac{\textrm{Cell body length}}{\textrm{Cell body width}}
\end{eqnarray*}
so that large (small) $R_s$ corresponds to slender (rounded) bodies.  The
relations of mean velocity, mean power and performance to the ratio of the cell
body sizes $R_s$ are given in Figure~$\ref{fig.sc.21}$, from which we see that
slender cells swim faster than rounded ones (Figure~$\ref{fig.sc.21}$(a)), yet
they require more power expenditure (Figure~$\ref{fig.sc.21}$(b)) and the
performance is worse (Figure~$\ref{fig.sc.21}$(c)). We should note that although
in general  Barry's cell is more slender than van Haastert's cell, yet Barry's cell is much
more smaller, thus it swims more slowly,  yet with better performance
(Figure~$\ref{fig.sc.19}$). 

Based on our observations on the protrusion height and the cell's slenderness, we conclude that
\begin{enumerate} 

\item Within a reasonable range, protrusions with large
height will result in faster swimming and better performance.  

\item Slender
cells swim faster, but their performance is worse than those rounded ones.
\end{enumerate}

\subsection{Do  asymmetric  shape deformations improve swimming?} 
 \label{asymm}

In Sections~$\ref{Ddsim} - \ref{Height}$ we discussed symmetric shape
deformations for amoebae swimming at LRN, but to date there is no clear evidence
which shows that amoebae favor such symmetric modes . Rather, they seem to
prefer asymmetric modes in which the protrusions travel one by one
(Figure~$\ref{fig.sc.3}$) \citep{VanHaastert:2011:ACU,Barry:2010:DAN}. In the 2D
model of swimming (Figure~$\ref{fig.sc.3}$(b)) this means that the protrusions
at the two sides alternate rather than appearing symmetrically, so that the cell
swims in a snake-like trajectory.  This raises the question as to why amoebae
employ the asymmetric mode -- are there advantages over the symmetric mode or
are protrusions constrained by the internal dynamics?

\begin{figure}[htbp]
\centering
\includegraphics[width=1\textwidth]{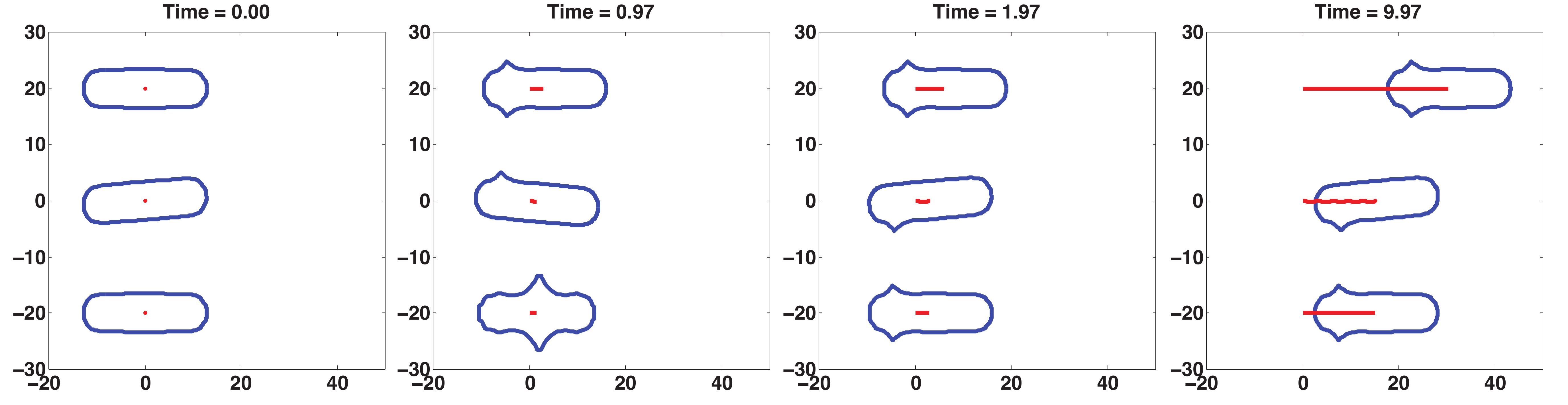}
\caption{A comparison of the three swimming amoebae. The top and bottom rows
use the same sequence of shape changes, but protrusions in the bottom row
travel at half the speed of those in the top row. The asymmetric swimmer in the
center row alternates the protrusions on the two sides. The period for the
symmetric swimmers is 1 min, while for the asymmetric swimmer it is 2 min. The axes are
in units of  $\mu m$.}
\label{fig.Comp3Cell}
\end{figure}

We design an asymmetric swimmer for which the protrusions alternate sides during
successive cycles, and adapt the previous numerical scheme.  Asymmetric
swimmers are not rotation-free, hence we must consider the torque on the
swimmer, and the rotational velocity is calculated via
 ($\ref{eq.review2D.27}$) (Figure~$\ref{fig.Comp3Cell}$, center). We
compare the asymmetric swimmer with two symmetric swimmers whose cell body and
protrusion shapes are identical, and whose conformal mapping is truncated at the
same order $N$. One of them propagates its pair of protrusions at the same speed
as the asymmetric swimmer (Figure~$\ref{fig.Comp3Cell}$, top), while the other
propagates its pair of protrusions at half the speed of the others
(Figure~$\ref{fig.Comp3Cell}$, bottom).
The red lines in Figure~$\ref{fig.Comp3Cell}$ reflect the trajectories of the
center of mass of each swimmer, and one sees that both symmetric swimmers move
in a straight line (i.e., rotation-free) while the asymmetric swimmer in the
center images swims in a slightly snake-like style. The simulations show that
the symmetric swimmer propagating its pair of protrusions at the same speed
(Figure~$\ref{fig.Comp3Cell}$, above) as the asymmetric swimmer travels a
distance of $X \sim 30.33$ for $T = 10$, while the asymmetric swimmer travels a
distance of $X \sim 15.36$ in the same time, which is slightly more than half
the distance traveled by the symmetric swimmer and slightly more than that for
the symmetric swimmer in (Figure~$\ref{fig.Comp3Cell}$, below).  The average
power of the asymmetric swimmer is again almost half that of the symmetric one
with the same protrusion speed, hence their performance is the same.  Thus we
find that rotation that results from the asymmetric shape deformations does not
lead to a reduction of performance, since the swimmer expends half the energy in
swimming half as fast.

\begin{figure}[htbp]
\centering
\includegraphics[width=1\textwidth]{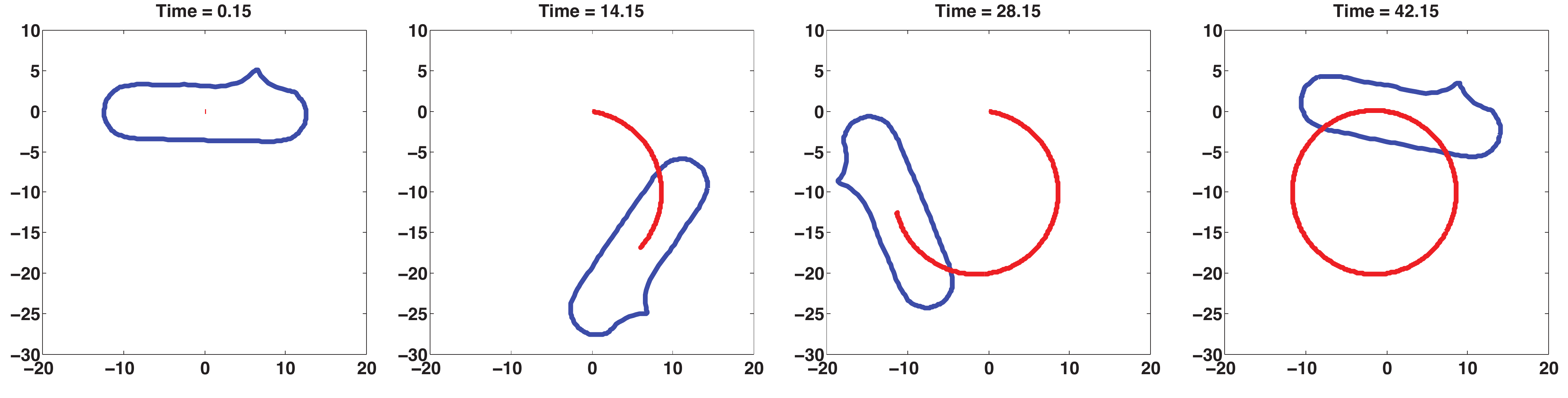}
\caption{The rotating amoeba. Lengths are measured in microns.}
\label{fig.RotAmoebae}
\end{figure}

Finally we show that when a protrusion always moves along one side, the cell
simply rotates in the long run. In this case the global trajectory of the
center of mass is a circle generated after a sufficient number of cycles
(Figure~$\ref{fig.RotAmoebae}$).

 \section{Discussion}
\label{Sec.Dis}

Movement of eukaryotic cells is best-characterized for keratocytes, which use
actin-driven lamellipodia at the leading and myosin-driven contraction at the
rear to move \citep{Keren:2008:BAA}. They move with very little shape change,
which simplifies their description, and several models that reproduce the shape
have been proposed\citep{Herant:2010:FFC,Rubinstein:2009:AMV,Wolgemuth:2011:RMS}.
Understanding of the mechanical balances that produce stable gliding motion is
emerging, but the linkage between the biochemical state and the mechanical state
is still not understood.  For example, how an initially symmetric cell breaks
the symmetry of the rest state and begins to move, and how the localization of
the control molecules and the properties of substrate affect motion, has not
been explained.

In contrast, other cells are found to use very complicated shape changes for
 locomotion, and this has led to the overarching question posed by
 experimentalists, which is `How does
 deformation of the cell body translate into
 locomotion?'\citep{Renkawitz:2010:MFG}.  The model and results described herein
 on swimming by shape changes was motivated by recent experiments which show
 that both neutrophils and Dd can swim -- in the strict sense of propelling
 themselves through a fluid without using any attachments -- in response to
 chemotactic gradients.  Our results for 2D cells show how the shape and height
 of a protrusion affect the speed and efficiency of swimming and may give
 insights into optimal designs of  micro-robots.

The protrusions and other shape changes used in swimming require forces that
 must be correctly orchestrated in space and time to produce net motion, and to
 understand this orchestration one must couple the cellular dynamics with the
 dynamics of the surrounding fluid or ECM.  This remains an open problem for
 future research.

\begin{acknowledgements}
 Supported in part by NSF Grant DMS \#s 9517884 and 131974 to H. G. Othmer. 
\end{acknowledgements}

 \appendix
\appendixpage
\addappheadtotoc

\section{Translation of the swimmer: proof of  (\ref{eq.review2D.24})}\label{Appendix:A_tr}

\begin{proof}
 $U_\infty$ is
defined as the translation part of the velocity field at infinity
 \citep{Shapere:1989:GSP},
\begin{eqnarray*}
U_\infty = \lim_{R \rightarrow \infty} \oint \dfrac{d \theta}{2 \pi} u 
\end{eqnarray*}
where $u$ is the fluid velocity field in the exterior domain.
In 2D it can be expressed by the complex integral
\begin{eqnarray}\label{eq.review2D.23}
U_\infty = \lim_{R \rightarrow \infty} \dfrac{1}{2 \pi i} \oint_{|z| = R} \dfrac{u (z, \overline{z})}{z} dz  
\end{eqnarray}
Choose $R > 0$ large enough so that $\phi$ and  $\chi'$ are analytic on $|z| >
 R$ and  continuous on  $|z| \geq R$. Let $\xi = z/ R$ and then  by  ($\ref{eq.review2D.23}$) and Table~$\ref{tab.1}$,
\begin{eqnarray}
\label{eq.C.5}
  U_\infty  = \lim_{R \rightarrow \infty} \dfrac{1}{2 \pi i} \oint_{|\xi|
 = 1} \Big[ \dfrac{\phi (R \xi)}{\xi} -R  \overline{\phi' (R \xi)}  
- \dfrac{\overline{\chi' (R \xi)}}{\xi} \Big] d \xi. 
\end{eqnarray}
Let 
\begin{eqnarray*}
 \widetilde{\phi} (\xi) = \phi' (R \xi), \qquad \widetilde{\psi} (\xi) = \chi' (R \xi)
\end{eqnarray*}
Obviously, $\widetilde{\phi}$, $\widetilde{\psi}$ are analytic on $|\xi|
 > 1$ (including at infinity) and  continuous on $|\xi| \geq 1$. Define 
\begin{eqnarray*}
 \overline{\widetilde{\phi}} (\xi) = \overline{ \widetilde{\phi} (\overline{\xi})} \qquad \textrm{and}  \qquad
\overline{\widetilde{\psi}} (\xi) = \overline{ \widetilde{\psi} (\overline{\xi})} 
\end{eqnarray*}
Then as functions of $\xi$, $\overline{\widetilde{\phi}} (1/\xi)$ and
 $\overline{\widetilde{\psi}} (1/\xi)$ are analytic on $|\xi| < 1$ and  
continuous on $|\xi| \leq 1$, and on $|\xi| = 1$ we have
\begin{eqnarray*}
 \overline{\widetilde{\phi}} \big( \dfrac{1}{\xi} \big) = \overline{\widetilde{\phi} (\xi)}, \qquad
 \overline{\widetilde{\psi}} \big( \dfrac{1}{\xi} \big) = \overline{\widetilde{\psi} (\xi)}
\end{eqnarray*}
Therefore  by the Cauchy Integral Theorem 
\begin{eqnarray*}
 \dfrac{1}{2 \pi i} \oint_{|\xi| = 1} \overline{\phi' (R \xi)} d \xi &=&
 \dfrac{1}{2 \pi i} \oint_{|\xi| = 1} \overline{\widetilde{\phi}} \big(
 \dfrac{1}{\xi} \big) d \xi = 0    \\
 \dfrac{1}{2 \pi i} \oint_{|\xi| = 1}  \dfrac{\overline{\chi' (R \xi)}}{\xi} d \xi &=&  \dfrac{1}{2 \pi i} \oint_{|\xi| = 1} \dfrac{1}{\xi} 
 \overline{\widetilde{\psi}} \big( \dfrac{1}{\xi} \big) d \xi =  \overline{\widetilde{\psi}} \big( \dfrac{1}{\xi} \big) \Big|_{\xi = 0} =\overline{\chi' (\infty)} = \overline{b_0}.
\end{eqnarray*} 
For the first term in \ref{eq.C.5}, it follows by use of the Residue Theorem that 
\begin{eqnarray*}
 \dfrac{1}{2 \pi i} \oint_{|\xi| = 1}  \dfrac{\phi (R \xi)}{\xi} d \xi = a_0.
\end{eqnarray*}
Moreover, since the conformal mapping $z = w (\zeta) $ has the form
  ($\ref{eq.review2D.13}$), it is easily seen that $a_0 = A_0$ and $b_0
 = B_0$.  
Hence we have proven the assertion in   ($\ref{eq.review2D.24}$).

\end{proof}

\section{Rotation of the swimmer: proof of  (\ref{eq.review2D.27})}
\label{Appendix:A_rot}

Suppose that the swimmer has the current shape and velocity field given by
$z = w(\zeta)$ and $V(\sigma)$, respectively, and  denote the resulting torque by $T (V; w)$.
The resulting rotational velocity $\omega $ can be calculated by considering a uniform rotation
of a rigid swimmer with the same shape $z = w(\zeta)$ and resulting in the same torque. Such a uniform 
rotational velocity field can be expressed as
\begin{eqnarray*} 
 \widetilde{V} (\sigma  ) = i  \omega w (\sigma )
\end{eqnarray*}
Let 
\begin{eqnarray*} 
 V^{\textrm{rot}} (\sigma ; t) = i   w (\sigma;t)
\end{eqnarray*}
i.e., $ \widetilde{V} =\omega V^{\textrm{rot}}$. To match the torque
that results from the two velocity 
fields, we have $T (V; w) = T (\omega V^{\textrm{rot}}; w)$.
Thus $\omega $ can be expressed as
\begin{eqnarray*} 
\omega = \dfrac{T (V ; w)}{T (V^{\textrm{rot}}; w)}
\end{eqnarray*}
which gives the first relation in  (\ref{eq.review2D.27}). 

For the other two relations in  (\ref{eq.review2D.27}), first we show
that $b_{-1} = B_{-1} \alpha_{-1}$ as follows.  From
 ($\ref{eq.review2D.26}$) it is easily seen that 
\begin{eqnarray*}
 b_{-1} = \dfrac{1}{2 \pi i } \int_{\partial\Omega} \psi (z ) dz. 
\end{eqnarray*}
Since $\partial \Omega = \{ w (\sigma; t); \sigma \in S^1 \}$, the above equation can be transformed into
\begin{eqnarray}\label{eq.review2D.30}
  b_{-1} = \dfrac{1}{2 \pi i } \int_{S^1} \Psi (\sigma) w' (\sigma) d \sigma.
\end{eqnarray}
Finally from equations~($\ref{eq.review2D.13}$,$\ref{eq.review2D.16}$,$\ref{eq.review2D.30}$) we have
\begin{eqnarray*}
   b_{-1} = \dfrac{1}{2 \pi i } \int_{S^1} 
\big( B_0 + \dfrac{B_{-1}}{\sigma} + \dfrac{B_{-2}}{\sigma^2} + \cdots \big)
\big( \alpha_{ 1} - \dfrac{\alpha_{-1}}{\sigma^2} - \dfrac{2 \alpha_{-2}}{\sigma^3} - \cdots \big) d \sigma
= B_{-1} \alpha_{ 1}
\end{eqnarray*}

Finally we prove the relation $T (V ; w) =- 4 \pi \mu \Im b_{-1} $ as follows.

\begin{proof}

The torque associated to the boundary condition $V (\sigma)$ is given by
\begin{eqnarray*}
 T (V;w) = \lim_{R \rightarrow \infty} \Im \oint r \times f ds 
= \lim_{R \rightarrow \infty} \Im \Big[ \int_{|z| = R} \overline{z} f ds \Big]
\end{eqnarray*}
From Table~$\ref{tab.1}$ we see that $f$ is a sum of two parts: $4 \mu ( \Re \phi' ) n$ along the $n$ direction and 
$- 2 \mu (z \overline{\phi''} + \overline{\chi''}) \overline{n}$ along the $\overline{n}$ direction, where
$n = - i d z / d s$ is the exterior normal on $\partial\Omega$. When taking the cross product with $r$, the
first part necessarily vanishes since it is parallel to $r$. 
Hence
\begin{eqnarray*}
 \oint r \times f ds = - 2 i \mu \int_{|z| = R} \overline{z} \Big( z \overline{\phi''} + \overline{\chi''} \Big) d \overline{z}
\end{eqnarray*}
and
\begin{eqnarray*}
 T ( V ; w ) = - 2 \mu \lim_{R \rightarrow \infty} \Re  \int_{|z| = R} \overline{z} \Big( z \overline{\phi''} + \overline{\chi''} \Big) d \overline{z}
\end{eqnarray*}
On the other hand we have
\begin{eqnarray*}
 \dfrac{\partial u}{\partial \overline{z}} = - \big( z \overline{\phi''} + \overline{\chi''} \big)
\end{eqnarray*}
so
\begin{eqnarray*}
 T (V ; w) &=& 2 \mu \lim_{R \rightarrow \infty} \Re  \int_{|z| = R}
 \overline{z} \dfrac{\partial u}{\partial \overline{z}}   d \overline{z} 
= 2 \mu \lim_{R \rightarrow \infty} \Re \Big[ -  \int_{|z| = R} u d \overline{z}  \Big]   \\
&=& 4 \pi \mu  \lim_{R \rightarrow \infty}  \Im \Big[ \dfrac{1}{2 \pi i}
\int_{|z| = R} u (z , \overline{z}) d \overline{z} \Big]. 
 \end{eqnarray*}

Now we only need to calculate the complex integral in the above equation. We
proceed as  in the proof for $U_\infty$. For $R$ large enough,
we introduce the substitution $\xi = z / R$ and then we have
\begin{eqnarray*}
  \dfrac{T (V ; w)}{4 \pi \mu} = \lim_{R \rightarrow \infty} \Im \Big[  \dfrac{R}{2 \pi i} \oint_{|\xi| = 1} \Big( \phi (R \xi)
 - R \xi \overline{\phi' (R \xi)} - \overline{\chi' (R \xi)} \Big) d \overline{\xi} \Big]
\end{eqnarray*}
On $|\xi| = 1$, we have $d \overline{\xi} = d \xi^{-1} = - \xi^{-2} d \xi$, so
\begin{eqnarray*}
   \dfrac{T (V ; w)}{4 \pi \mu} = \lim_{R \rightarrow \infty} \Im \Big[ -
   \dfrac{R}{2 \pi i} \oint_{|\xi| = 1} \Big( \dfrac{\phi (R \xi)}{\xi^2} 
 - \dfrac{R}{\xi}  \overline{\phi' (R \xi)} - \dfrac{1}{\xi^2} \overline{\chi' (R \xi)} \Big) d \xi \Big].
\end{eqnarray*}
By the Residue Theorem
\begin{eqnarray*}
 \dfrac{1}{2 \pi i} \oint_{|\xi| = 1} \dfrac{\phi (R \xi)}{\xi^2} d \xi = 0
\end{eqnarray*}
and by the Cauchy Integral Theorem
\begin{eqnarray*}
 \dfrac{1}{2 \pi i} \oint_{|\xi| = 1} \dfrac{1}{\xi}  \overline{\phi' (R \xi)}
 d \xi =  \dfrac{1}{2 \pi i} \oint_{|\xi| = 1} \dfrac{1}{\xi}  
\overline{\widetilde{\phi}} (\dfrac{1}{\xi}) d \xi = \overline{\widetilde{\phi}}
(\dfrac{1}{\xi}) \Big|_{\xi = 0} = \overline{\phi' (\infty)} = 0 
\end{eqnarray*}
where $\overline{\phi' (\infty)} = 0$ because of the form of $\phi$ given in
 ($\ref{eq.review2D.25}$). Thus the first two terms vanish, and by the
Cauchy  Integral Theorem,  
\begin{eqnarray*}
  \dfrac{1}{2 \pi i} \oint_{|\xi| = 1} \dfrac{1}{\xi^2} \overline{\chi' (R \xi)} d \xi =  
  \dfrac{1}{2 \pi i} \oint_{|\xi| = 1} \dfrac{1}{\xi^2} \overline{\widetilde{\psi}} (\dfrac{1}{\xi}) d \xi
 = \dfrac{d}{d \xi} \Big[ \overline{\widetilde{\psi}} (\dfrac{1}{\xi}) \Big] \Big|_{\xi = 0} 
= \dfrac{d}{d \xi} \Big[ \overline{\chi' \big( \dfrac{R}{\overline{\xi}} \big)} \Big] \Big|_{\xi = 0} 
\end{eqnarray*}
By  ($\ref{eq.review2D.26}$), on $|\xi| \leq 1$ we have
\begin{eqnarray*}
  \overline{\chi' \Big( \dfrac{R}{\overline{\xi}} \Big)} = \overline{b_0} + 
  \dfrac{\overline{b_{-1}}}{R} \xi + \dfrac{\overline{b_{-2}}}{R^2} \xi^2 +
  \cdots +  \dfrac{\overline{b_{-n}}}{R^n} \xi^n + \cdots  
\end{eqnarray*}
thus 
\begin{eqnarray*}
  \dfrac{d}{d \xi} \Big[ \overline{\chi' \big( \dfrac{R}{\overline{\xi}} \big)}
  \Big] \Big|_{\xi = 0} = \dfrac{\overline{b_{-1}}}{R}. 
\end{eqnarray*}
Finally we have that 
\begin{eqnarray*}
  \dfrac{T (V ; w)}{4 \pi \mu} = \Im \Big[   R \dfrac{\overline{b_{-1}}}{R} \Big] = - \Im b_{-1}
\end{eqnarray*}

\end{proof}

\section{Algorithm for the shape changes}
\label{Appendix:Pseudo_Code}
The “inner skeleton”, {\em i.e.}  the red contour in Fig 4.3, is a polygon of
cross-like shape, with a semi-circle  at each end of an arm. The polygon
in each step consists of 28 vertices, where each semi-circle end has 6
vertices. The initial shape p0 (i.e, shape at t=0) is generated as follows.
\medskip
\begin{itemize}
\item[]p0(1)=2+2i\\
p0(7)=17.5-2i\\
p0(8)=17.5-2.2i\\
p0(14)=19.5-2i\\
p0(15)=20-2i\\
p0(21)=19.5+2i\\
p0(22)=19.5+2.2i\\
p0(28)=17.5+2i\\
 \end{itemize}
for k=1:1:5
 \begin{itemize}
\item[]   p0(1+k) = 2+ 2i*exp(k*1i*pi/5)\\
    p0(8+k)=18.5-2.2i - 1*exp(k*1i*pi/5)\\
    p0(15+k)=20-2i*exp(k*1i*pi/5)\\
    p0(22+k) = 18.5 + 2.2i + 1*exp(k*1i*pi/5)\\
\end{itemize}
end
 
p0(29)=p0(1)\\

For each time step dt, move the two arms (i.e., the “blebs”) of the cross
forward simultaneously. We change the length of the blebs and the body so as to
simulate the grow and decay of the blebs, and compensate for the resulting area
change. Below is the pseudo-code that generates the inner skeleton p at the k$^{th}$
time step.  
\medskip

\noindent
for j=1:1:6 \\
\indent p(j) = p0(j) + 0.5*H*sin(dt*k*pi)\\ 
end

\smallskip
p(7)= p(7) - 0.5\\

\noindent
for j=8:1:13\\
  \indent p(j) = p0(j)- 0.5*k -  H*sin(dt*k*pi)*1i\\
end\\

p(14) = p(14) - 0.5\\

\noindent
for j=15:1:20\\
\indent  p(j) = p0(j) - 0.5*H*sin(dt*k*pi)\\  
end\\

p(21) = p(21) - 0.5\\

\noindent 
for j=22:1:27\\
\indent p(j) = p0(j) - 0.5*k +  H*sin(dt*k*pi)*1i\\
end\\

p(28) = p(28) - 0.5\\

p(29)=p(1)\\



\end{document}